\documentclass[english,aps,twocolumn]{revtex4}
\usepackage[T1]{fontenc}
\usepackage[latin9]{inputenc}
\usepackage{amsmath}
\usepackage{amssymb}
\usepackage{graphicx}

\makeatletter
\@ifundefined{textcolor}{}
{%
 \definecolor{BLACK}{gray}{0}
 \definecolor{WHITE}{gray}{1}
 \definecolor{RED}{rgb}{1,0,0}
 \definecolor{GREEN}{rgb}{0,1,0}
 \definecolor{BLUE}{rgb}{0,0,1}
 \definecolor{CYAN}{cmyk}{1,0,0,0}
 \definecolor{MAGENTA}{cmyk}{0,1,0,0}
 \definecolor{YELLOW}{cmyk}{0,0,1,0}
}

\makeatother

\usepackage{babel}
\begin{document}

\preprint{This line only printed with preprint option}

\title{Phases dynamics in VCSELs with delayed optical feedback and cross
re-injection }

\author{J. Javaloyes$^{(1)}$, M. Marconi$^{(2)}$ and M. Giudici$^{(2)}$}

\affiliation{$^{(1)}$ Departament de Fisica, Universitat de les Illes Baleares,
C/ Valldemossa km 7.5, 07122 Mallorca, Spain\\
 $^{(2)}$ Institut Non-Linéaire de Nice, Université de Nice Sophia
Antipolis, CNRS UMR 7335, 06560 Valbonne, France}
\begin{abstract}
We study theoretically the non linear polarization dynamics of Vertical-Cavity
Surface-Emitting Lasers in the presence of an external cavity providing
delayed optical feedback and cross polarization re-injection. We show
that far from the laser threshold, the dynamics remains confined close
to the equatorial plane of a Stokes sphere of a given radius and we
reduce the dynamics to a dynamical system composed of two phases:
the orientation phase of the quasi-linear polarization and the optical
phase of the field. We explore the complex modal structure given by
the double feedback configuration and recovers as particular cases
the Lang-Kobayashi modes and the modes founds by Giudici et al. \cite{MGJ-PRA-07}.
We also re-interpret the square waves switching dynamics as phase
kinks.
\end{abstract}
\maketitle

\section{introduction}

Vertical-cavity surface-emitting lasers (VCSELs) possess several advantages
compared to conventional semiconductor edge-emitting lasers. Their
circular aperture induces a high beam quality as compared to the strongly
astigmatic output of edge emitters. In addition, the possibility to
couple them efficiently to optical fiber as well as the possibility
to perform on-wafer testing renders them superior to classical laser
diodes.

VCSELs of large transverse dimensions can present rich spatio-temporal
transverse dynamics \cite{MBS-ISLC-00,MB-JQE-02,MB-PRA-02}, which
can be harnessed for instance in order to create transverse localized
structures\cite{BTB-NAT-02,GBG-PRL-10}. However, such complexity
can be avoided by choosing a VCSEL only a few micrometers wide thereby
limiting the number of transversal modes and since VCSELs are also
intrinsically single-longitudinal mode, one may consider them ideal
single mode devices. 

However, these devices exhibit a nearly degenerate polarization orientation
owing to their almost perfect symmetry around the cavity axis. Usually,
the two polarization modes are aligned along the $\left[1\,1\,0\right]$
and the $\left[1\,-1\:0\right]$ crystallographic axes, although some
randomness exists due to the presence of hardly controllable strain
\cite{vanDoorn}. In addition to the complex problem of the elasto-optic
effects \cite{VPV-JQE-06}, the application of a voltage to the laser
diode can also induce anisotropies via an electro-optic effect \cite{electrooptic}.
The existence of such favored directions is sufficient to weakly pin
the polarization orientation dynamics and to define two optical modes
having slightly different losses and frequencies. Such residual anisotropies
are termed dichroism and birefringence in the cavity. 

These two modes exhibit a high degree of coupling as they share an
identical transverse spatial profile and the same carrier reservoir
allowing for the existence of complex polarization dynamics. VCSELs
are prone to display polarization switching \cite{CRL-APL-94,BTS-OPL-99,AS-APL-01}
accompanied in some cases by regions of polarization bistability or
even regimes where the polarization of the output oscillate in time
\cite{SWA-PRA-03}.

In addition, the quasi-degeneracy of the orthogonal polarization states
in VCSELs enables efficient cross-gain modulation among orthogonal
polarizations when the device is used as an optical amplifier. In
particular, VCSELs have been proposed as promising devices for implementing
useful dynamics taking advantage of their polarization degree of freedom
\cite{PSG-AOT-11}. When VCSELs are subject to optical feedback, the
polarization stability is affected and polarization dynamics appear
even in the case of perfectly isotropic feedback \cite{GAB-JOSAB-99}.
Polarization-rotated optical feedback ---where the two linearly polarized
components, LP-x and LP-y, of the light emitted by the device are
fed back into the laser cavity after the LP-x component is converted
into the LP-y component and vice-versa--- induces a regular polarization
dynamics which can be as fast as 9 GHz \cite{RBC-JQE-97,LHG-APL-98}.
Such symmetrical cross re-injection was found to induce waveforms
ranging from square-waves to sinusoidal oscillations. 

Asymmetrical cross polarization re-injection (XPR), of a single polarization
into the orthogonal one, was also shown to promote the occurrence
of square wave switching \cite{MGJ-PRA-07} between orthogonal polarization
with a repetition period close to twice the delay taken by the light
to come back into the rotated orientation. The quasi-degeneracy of
the VCSEL allows finding rather easily this regime which also exists
in edge-emitting devices \cite{GES-OL-06} yet for higher threshold
of XPR. The analysis of laser system with multiple delays and different
type of optical feedback are however not widespread. A few years ago
some of the authors have proposed to combine XPR with polarization
selective optical feedback (PSF), in order to achieve passive mode-locking
in VCSELs \cite{JMB-PRL-06,MJB-JQE-07}. Recently, we also showed
that PSF can be used to tune and control the existence of the square
wave switching generated by XPR \cite{MJB-PRA-14}. However, VCSELs
must be described by a relatively high dimensional dynamical system
that would consider the dynamics of the two polarizations as well
as their interplay for the two carriers reservoirs with opposite spin
orientations. This fact, in addition to the presence of two different
time delays, render the analysis formidable a problem and a reduction
to a lower dimensional system as presented in \cite{EDP-PRA-99} for
a solitary VCSEL close to the laser threshold, would be highly beneficial.

In this manuscript we show that far from the laser threshold, the
dynamics of the VCSEL remains confined close to the equatorial plane
of a Stokes sphere of a given radius. This allows us to decouple the
relaxation oscillation for the total emitted power as well as the
fluctuations in the ellipticity of the emitted light. We reduce the
dynamics to a dynamical system composed of two phases: the orientation
phase of the quasi-linear polarization and the optical phase of the
field. 

We believe that such phase model and the general methodology employed
here can be useful to harness the phase and orientation dynamics of
VCSEL far from threshold, an important parameter regime for most applications.
Indeed, such reduction not only allows to simplify the analytical
and numerical studies but it may also be useful to get insight into
future applications. While optical information is usually encoded
in binary levels of light intensity, next generation communication
systems will process not only intensity but also phase and the polarization
data. Here, the simplicity of the phase model allowed us to explore
the complex modal structure given by the double feedback configuration
but also to re-interpret the square waves switching dynamics \cite{JMB-PRL-06,MJB-JQE-07,SGP-PRA-2012}
as polarization orientation kinks.

The manuscript is organized as follows. In section II we briefly recall
the basis of the model we use and fix the order of magnitude of the
parameters for which our analysis applies. Section III is devoted
to the phase reduction while we discuss the modal structure of our
model in section IV. Perspective and conclusions are drawn in section
V.

\section{The model}

We base our theoretical analysis on the so-called spin-flip model
(SFM) \cite{SFM-PRA-95}, suitably modified for incorporating the
effects of PSF and XPR. We assume that the LP-y mode is feedback into
itself (PSF) and cross re-injected into the LP-x mode (XPR). While
the most direct way to incorporate XPR and PSF would be in terms of
the linearly polarized components of the field, $X$ and $Y$, the
derivation of the phase model is more natural in the circular basis.
The SFM model expressed in circular component reads
\begin{eqnarray}
\dot{E}_{\pm} & \negthickspace=\negthickspace & \left(1+i\alpha\right)\left(G_{\pm}-1\right)E_{+}-zE_{\mp}+C_{\pm},\label{eq:Epm}\\
T\dot{D}_{\pm} & \negthickspace=\negthickspace & 1+P-D_{\pm}-G_{\pm}\left|E_{\pm}\right|^{2}\mp\gamma_{J}\left(D_{+}-D_{-}\right),\label{eq:Dpm}
\end{eqnarray}
where $E_{\pm}$ are the amplitudes of the circular left and right
components of the field and $D_{\pm}$ are the scaled carrier density
in the two spin channels. In the equations (\ref{eq:Epm}-\ref{eq:Dpm}),
time has been scaled to the cavity decay rate $\kappa$, while $T=\kappa/\gamma_{e}$
represents the scaled carrier lifetime and $\gamma_{s}$ is the spin-flip
and carrier relaxation rate normalized to $\gamma_{e}$. The rate
of carrier density injected into the active region above threshold
is represented by $P$. In addition, $\alpha$ stands for the linewidth
enhancement factor \cite{H-JQE-82} and the complex parameter $z=\gamma_{a}+i\gamma_{p}$
is composed of the linear dichroism $\gamma_{a}$ and the birefringence
$\gamma_{p}$. We introduced the effect of ultra-fast gain saturation
in the expression of the gain as 
\begin{eqnarray}
G_{\pm} & = & D_{\pm}\left(1-\frac{\varepsilon_{g}}{2}\left|E_{\pm}\right|^{2}\right)\label{eq:Gnl}
\end{eqnarray}
 with the parameter of self-saturation $\varepsilon_{g}$. Since each
component of the field only interacts with only one of the two spin
channels, direct cross saturation between $E_{+}$ and $E_{-}$ does
not exist. Several sources can contribute to the factor $\varepsilon_{g}$
like for instance spatial hole burning, spectral hole burning as well
as carrier heating. In the case where the $Y$ component is being
feedback with complex rate $\eta\exp\left(-i\Omega\right)$ after
a time $\tau_{f}$ and cross re-injected into the $X$ polarization
with a rate $\beta\exp\left(-ia\right)$, the expression of $C_{\pm}$
reads 
\begin{eqnarray}
C_{\pm} & = & \frac{\beta}{2i}e^{-ia}\left(E_{+}^{\tau_{r}}-E_{-}^{\tau_{r}}\right)\pm\frac{\eta}{2}e^{-i\Omega}\left(E_{+}^{\tau_{f}}-E_{-}^{\tau_{f}}\right).\label{eq:fb_xpr}
\end{eqnarray}

The modulus $M_{\pm}$ and phase $F_{\pm}$ decomposition of $C_{\pm}$
as defined as $C_{\pm}=M_{\pm}\exp\left(iF_{\pm}\right)$ is

\begin{widetext}

\begin{eqnarray}
M_{+} & = & \frac{\beta}{2}R_{+}^{\tau_{r}}\sin\left(\psi_{+}^{\tau_{r}}-\psi_{+}-a\right)-\frac{\beta}{2}R_{-}^{\tau_{r}}\sin\left(\psi_{-}^{\tau_{r}}-\psi_{+}-a\right)+\frac{\eta}{2}R_{+}^{\tau_{f}}\cos\left(\psi_{+}^{\tau_{f}}-\psi_{+}-\Omega\right)-\frac{\eta}{2}R_{-}^{\tau_{f}}\cos\left(\psi_{-}^{\tau_{f}}-\psi_{+}-\Omega\right),\\
M_{-} & = & \frac{\beta}{2}R_{+}^{\tau_{r}}\sin\left(\psi_{+}^{\tau_{r}}-\psi_{-}-a\right)-\frac{\beta}{2}R_{-}^{\tau_{r}}\sin\left(\psi_{-}^{\tau_{r}}-\psi_{-}-a\right)-\frac{\eta}{2}R_{+}^{\tau_{f}}\cos\left(\psi_{+}^{\tau_{f}}-\psi_{-}-\Omega\right)+\frac{\eta}{2}R_{-}^{\tau_{f}}\cos\left(\psi_{-}^{\tau_{f}}-\psi_{-}-\Omega\right),\nonumber \\
F_{+} & = & -\frac{\beta}{2}R_{+}^{\tau_{r}}\cos\left(\psi_{+}^{\tau_{r}}-\psi_{+}-a\right)+\frac{\beta}{2}R_{-}^{\tau_{r}}\cos\left(\psi_{-}^{\tau_{r}}-\psi_{+}-a\right)+\frac{\eta}{2}R_{+}^{\tau_{f}}\sin\left(\psi_{+}^{\tau_{f}}-\psi_{+}-\Omega\right)-\frac{\eta}{2}R_{-}^{\tau_{f}}\sin\left(\psi_{-}^{\tau_{f}}-\psi_{+}-\Omega\right),\nonumber \\
F_{-} & = & -\frac{\beta}{2}R_{+}^{\tau_{r}}\cos\left(\psi_{+}^{\tau_{r}}-\psi_{-}-a\right)+\frac{\beta}{2}R_{-}^{\tau_{r}}\cos\left(\psi_{-}^{\tau_{r}}-\psi_{-}-a\right)-\frac{\eta}{2}R_{+}^{\tau_{f}}\sin\left(\psi_{+}^{\tau_{f}}-\psi_{-}-\Omega\right)+\frac{\eta}{2}R_{-}^{\tau_{f}}\sin\left(\psi_{-}^{\tau_{f}}-\psi_{-}-\Omega\right).\nonumber 
\end{eqnarray}

\end{widetext}

\subsection{Parameters range}

The polarization switchings in the SFM have been exhaustively analyzed
in the literature \cite{MPS-JQE-97,EDP-PRA-99}; Here, we consider
the case where the VCSEL may displays a large ranges of bistability
close to threshold and as such we consider small dichroism and birefringence
typically of the order of a few GHz. Typically, we take $\gamma_{p}=+5.24\times10^{-2}$,
which means that we denote by LP-x the reddest mode and $\gamma_{a}=0$.
Besides, we assume standard values for the Henry's factor, $\alpha=2$,
a normalized carrier lifetime $T=500$, and a typical normalized spin-flip
rate $\gamma_{s}=75$. The other typical values of the parameters
are $\eta\sim0.05$, $\beta\sim0.05$, $\varepsilon_{g}=0.02$ and
$P=10$ while the variance of the Gaussian white noise is $2\times10^{-2}$.

The presence of two kinds of feedback with possibly dissimilar delays
render the analysis of Eqs.~(\ref{eq:Epm}-\ref{eq:Dpm}) formidable
a problem. However, far from threshold one expects the dynamics that
involves the relaxations oscillations between the total emitted power
and the carrier reservoir to play only a minor role. Hence, in the
following of this manuscript we will assume large bias current is
$P\gtrsim10$, which corresponds typically to relaxation oscillations
of the order of $10\sim15\,$GHz. In realistic situations the strong
damping of the relaxations oscillations renders the laser almost a
Class-A system, although such definition is meaningful only for a
mono-mode system. Here, the standard ``unsaturated'' rate equations
and the SFM do not reproduce fairly this regime of strong damping
which explains why we included in our analysis the non linear saturation
in Eq.~(\ref{eq:Gnl}). Noteworthy, a completely identical phase
reduction as the one we discuss in this manuscript is possible without
relying on gain saturation, yet for unrealistic parameter ranges,
i.e. $P\sim100$, corresponding to a device biased one hundred time
above threshold. Several physical effects can contribute to the gain
compression parameter $\varepsilon_{g}$ like for instance spatial
hole burning in the transverse plane of the VCSEL, spectral hole burning
due to saturation of the individual intra-band transitions as well
as carrier heating. The latter has the effect to spread the carriers
within the band and to depopulate the available gain at the emission
wavelength. In the following, we assumed that $\varepsilon_{g}\in\mathbb{R}$
which is consistent with a situation dominated by spatial hole burning,
i.e. inter-band saturation.

\section{Phase reduction}

Far from threshold, the fluctuations of the total intensity will die
out rapidly and the dynamics will be confined on a Stokes sphere of
a given radius. Without external perturbation, one may not expect
any complex residual dynamics since this center manifold is only two
dimensional and consists in the polarization angle $\Phi$ and the
ellipticity parameter $\theta$, the optical phase being decoupled
from the rest. In addition, strongly elliptical states would incur
a large energetic penalty due to the imbalance between the two carrier
reservoirs further confining the residual dynamics to the vicinity
of the equatorial plane of the Stokes sphere and to polarization re-orientation.

Notwithstanding, the coherent delayed retro-actions imposed by the
feedback terms in Eq.~(\ref{eq:fb_xpr}) the optical phase of the
field couple back into the dynamics and as such our reduced model
will consist in a ``vectorial'' phase for the orientation of the
quasi-linear polarization and for the optical phase of the field.
It is worthwhile to notice that these two phases are of very different
nature. While the optical phase precise value is irrelevant due to
the phase invariance in an autonomous system the orientation phase
is not because of the pinning imposed by the dichroism and the birefringence.

\begin{figure}[tp]
\begin{centering}
\includegraphics[width=0.35\textwidth]{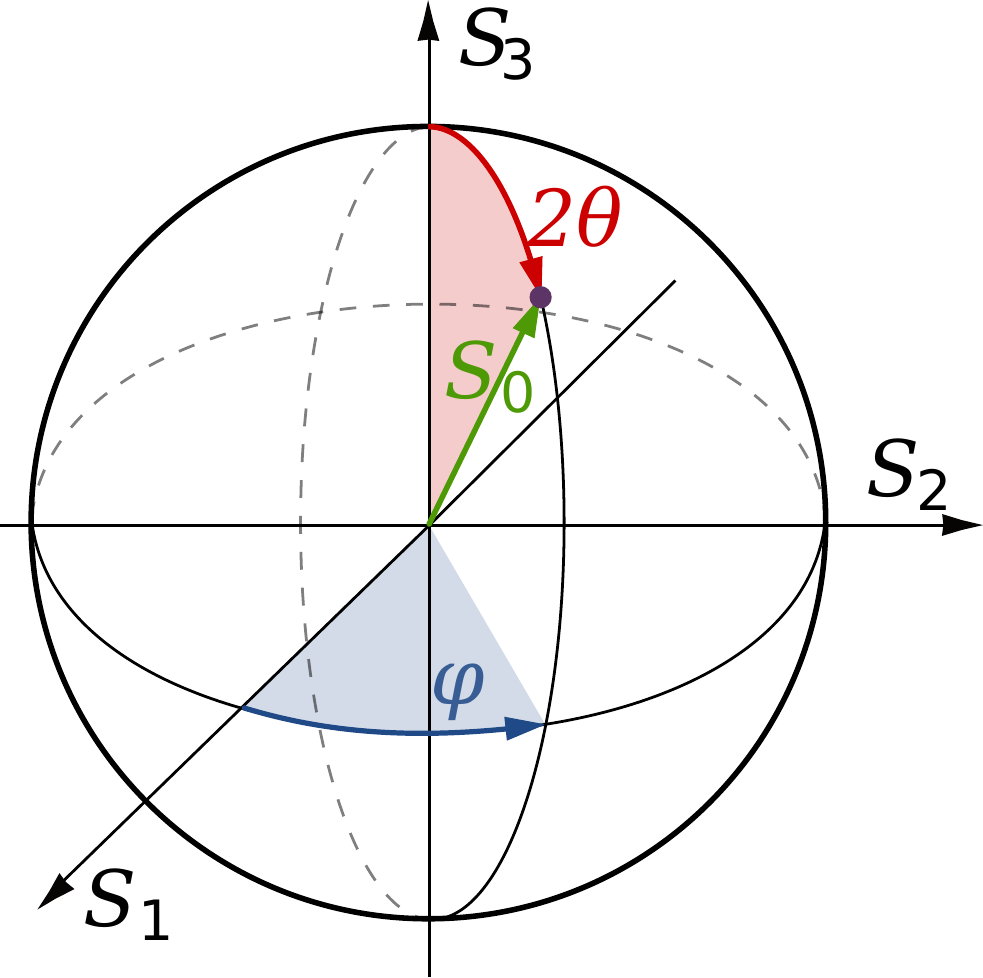} 
\par\end{centering}

\centering{}\caption{(Color Online): Angular representation of the VCSEL dynamics onto
the Stokes sphere.\label{fig:theo_0} }
\end{figure}

We now detail how the SFM with optical feedback and cross re-injection
can be reduced to such phase model far from threshold. We start by
separating the modulus and phase of the circular components by defining
$E_{\pm}=R_{\pm}\sqrt{2}\exp\left(i\psi_{\pm}\right)$, which yield
with $\Phi=\psi_{+}-\psi_{-}$ 
\begin{eqnarray}
\dot{R}_{+} & = & \left(N+n-1\right)R_{+}-\gamma_{a}R_{-}\cos\Phi-\gamma_{p}R_{-}\sin\Phi\label{eq:Rp}\\
 &  & -\varepsilon_{g}\left(N+n\right)R_{+}^{3}+M_{+},\\
\dot{R}_{-} & = & \left(N-n-1\right)R_{-}-\gamma_{a}R_{+}\cos\Phi+\gamma_{p}R_{+}\sin\Phi\nonumber \\
 &  & -\varepsilon_{g}\left(N-n\right)R_{-}^{3}+M_{-},\label{eq:Rm}\\
\dot{\psi}_{+} & = & \alpha\left(N+n-1\right)+\gamma_{a}\frac{R_{-}}{R_{+}}\sin\Phi-\gamma_{p}\frac{R_{-}}{R_{+}}\cos\Phi\nonumber \\
 &  & -\alpha\varepsilon_{g}\left(N+n\right)R_{+}^{2}+\frac{F_{_{+}}}{R_{+}},\label{eq:psip}\\
\dot{\psi}_{-} & = & \alpha\left(N-n-1\right)-\gamma_{a}\frac{R_{+}}{R_{-}}\sin\Phi-\gamma_{p}\frac{R_{+}}{R_{-}}\cos\Phi\nonumber \\
 &  & -\alpha\varepsilon_{g}\left(N-n\right)R_{-}^{2}+\frac{F_{-}}{R_{-}},\\
T\dot{N} & = & 1+P-N-\left(N+n\right)R_{+}^{2}-\left(N-n\right)R_{-}^{2}\nonumber \\
 &  & +\varepsilon_{g}\left[\left(N+n\right)R_{+}^{4}+\left(N-n\right)R_{-}^{4}\right],\label{eq:N}\\
T\dot{n} & = & -\gamma_{s}n-\left(N+n\right)R_{+}^{2}+\left(N-n\right)R_{-}^{2}\nonumber \\
 &  & +\varepsilon_{g}\left[\left(N+n\right)R_{+}^{4}-\left(N-n\right)R_{-}^{4}\right],\label{eq:n}
\end{eqnarray}
where we defined the average carrier density an $N=\left(N_{+}+N_{-}\right)/2$
and the imbalance between the two channels $n=\left(N_{+}-N_{-}\right)/2$
as well as $\gamma_{s}=1+2\gamma_{j}$. As previously mentioned, far
from threshold we expect the dynamics of the field to be restricted
to the vicinity of a Stokes sphere. We make that apparent by defining
the radius $I=R_{+}^{2}+R_{-}^{2}$ and the ratio of the two circular
components which is a measure the degree of ellipticity $\theta=\arctan\left(R_{-}/R_{+}\right)$,
hence $R_{+}=\sqrt{I}\cos\theta$ and $R_{-}=\sqrt{I}\sin\theta$.
We can relate the left (resp. right) circular components $E_{-}$
(resp. $E_{+}$) in terms of the Stokes coordinates $\left(S_{0},S_{1},S_{2},S_{3}\right)$
as described in Fig.~(\ref{fig:theo_0}) 
\begin{eqnarray}
S_{0} & = & \left|E_{-}\right|^{2}+\left|E_{+}\right|^{2}=2I,\\
S_{1} & = & 2\Re\left(E_{-}^{\star}E_{+}\right)=2I\sin\left(2\theta\right)\cos\Phi,\\
S_{2} & = & -2\Im\left(E_{-}^{\star}E_{+}\right)=2I\sin\left(2\theta\right)\sin\Phi,\\
S_{3} & = & \left|E_{-}\right|^{2}-\left|E_{+}\right|^{2}=2I\cos\left(2\theta\right).
\end{eqnarray}

In addition we proceed to the scaling of Eqs.~(\ref{eq:Rp}-\ref{eq:n})
to the natural time scale of the relaxation oscillation frequency
$\omega_{r}$ as $s=\omega_{r}t$ and define
\begin{equation}
\omega_{r}=\sqrt{\frac{2P}{T}}\:,\: D=\frac{2\left(N-1\right)}{\omega_{r}}\:,\: d=\frac{2n}{\omega_{r}}\:,\:\Gamma=\frac{\omega_{r}}{2}\left(1+P^{-1}\right)
\end{equation}
as well as the total intensity relatively to the steady value of the
solitary laser $S=I/P$. With our typical values for the parameters
far from threshold $P\sim10$ and $\omega_{r}\sim0.2$. As such $\Gamma\sim0.1$
and the oscillations are only mildly damped, i.e. the laser performs
ten oscillations before reaching its steady state in complete disagreement
with any experimental evidence. Here, the presence of gain saturation
strongly contributes to reduce the number of oscillation necessary
to reach the equilibrium. Upon simplification of Eqs.~(\ref{eq:Rp}-\ref{eq:n})
we get 

\begin{eqnarray}
\frac{dS}{ds} & = & DS+dS\cos\left(2\theta\right)-2\frac{\gamma_{a}}{\omega_{r}}S\sin\left(2\theta\right)\cos\Phi\nonumber \\
 & - & 2\frac{\varepsilon_{g}P}{\omega_{r}}S^{2}\mathcal{A}\left(S,\theta\right)+\frac{\mathcal{F}\left(S,\theta\right)}{\omega_{r}}\label{eq:S}\\
\frac{d\theta}{ds} & = & -\frac{d}{2}\sin\left(2\theta\right)-\frac{\gamma_{a}}{\omega_{r}}\cos\left(2\theta\right)\cos\Phi+\frac{\gamma_{p}}{\omega_{r}}\sin\Phi\nonumber \\
 & + & \frac{\varepsilon_{g}P}{4\omega_{r}}S\mathcal{B}\left(S,\theta\right)+\frac{\mathcal{G}\left(S,\theta\right)}{\omega_{r}}\label{eq:theta}\\
\frac{d\psi_{+}}{ds} & = & \frac{\alpha}{2}\left(D+d\right)+\frac{\gamma_{a}}{\omega_{r}}\tan\theta\sin\Phi-\frac{\gamma_{p}}{\omega_{r}}\tan\theta\cos\Phi\nonumber \\
 & - & \alpha\frac{\varepsilon_{g}P}{\omega_{r}}\mathcal{H}_{+}\left(S,\theta\right)+\frac{F_{_{+}}}{\omega_{r}R_{+}}\label{eq:psip2}\\
\frac{d\psi_{-}}{ds} & = & \frac{\alpha}{2}\left(D-d\right)-\frac{\gamma_{a}}{\omega_{r}}\mathrm{cotan}\theta\sin\Phi-\frac{\gamma_{p}}{\omega_{r}}\mathrm{cotan}\theta\cos\Phi\nonumber \\
 & - & \alpha\frac{\varepsilon_{g}P}{\omega_{r}}\mathcal{H}_{-}\left(S,\theta\right)+\frac{F_{-}}{\omega_{r}R_{-}}\label{eq:psim2}\\
\frac{dD}{ds} & = & -\Gamma D-\left(1+\frac{\omega_{r}D}{2}\right)\left(S-1\right)\label{eq:D}\\
 & - & \frac{\omega_{r}dS}{2}\cos\left(2\theta\right)+\frac{\varepsilon_{g}P}{2}S^{2}+\cdots\\
\frac{dd}{ds} & = & -\frac{\omega_{r}}{2}d\left(\frac{\gamma_{s}}{P}+S\right)-\left(1+\frac{\omega_{r}}{2}D\right)S\cos\left(2\theta\right),\label{eq:d}
\end{eqnarray}

In the equations~(\ref{eq:S}-\ref{eq:psim2}), we separated on the
first and second lines the natural contributions of the SFM model
from the ones due to feedbacks and non linear gain compression. The
definitions of $\mathcal{A},\mathcal{B},\mathcal{H}_{\pm},\mathcal{F}$
and $\mathcal{G}$ are 
\begin{eqnarray}
\mathcal{A}\left(S,\theta\right) & = & \left[1+\frac{\omega_{r}}{2}\left(D+d\right)\right]\cos^{4}\theta\nonumber \\
 & + & \left[1+\frac{\omega_{r}}{2}\left(D-d\right)\right]\sin^{4}\theta\\
\mathcal{B}\left(S,\theta\right) & = & \left(1+\frac{\omega_{r}}{2}D\right)\sin\left(4\theta\right)+\omega_{r}d\sin\left(2\theta\right)\\
\mathcal{H}_{+}\left(S,\theta\right) & = & \left(1+\frac{\omega_{r}}{2}\left(D+d\right)\right)S\cos^{2}\theta\\
\mathcal{H}_{-}\left(S,\theta\right) & = & \left(1+\frac{\omega_{r}}{2}\left(D-d\right)\right)S\sin^{2}\theta\\
\mathcal{F}\left(S\right) & = & 2\frac{R_{+}M_{+}+R_{-}M_{-}}{P}\\
\mathcal{G}\left(S\right) & = & \frac{M_{-}R_{+}-M_{+}R_{-}}{SP}
\end{eqnarray}
In addition, we neglected several terms in the carrier equations Eq.~(\ref{eq:D})
and Eq.~(\ref{eq:d}) which are due to non linear saturation. These
terms are of order $\mathcal{O}\left(\varepsilon_{g}P\omega_{r}\right)$
and are immaterial to our analysis. The dominant effect of gain compression
is to create an additional damping in the field equation leaving the
carrier dynamics essentially unchanged up to second order.

The equations~(\ref{eq:S}-\ref{eq:d}) can be further simplified
if one assume the following scaling of the parameters. We define $\varepsilon\sim0.2$
as our smallness parameter such that $\omega_{r}\sim\varepsilon$
and we consider the case where $\eta$ and $\beta$ as well as $\gamma_{a}$
and $\gamma_{p}$ are of order $\varepsilon^{2}$. Importantly, we
assume that the gain compression coefficient $\varepsilon_{g}$ is
also of order $\varepsilon^{2}$ although since $P\sim10$, the contribution
$\varepsilon_{g}\omega_{r}^{-1}P$ is considered to be of order one.
At last we assume that the spin-flip decay term scales like $\gamma_{s}\sim\varepsilon^{-2}$.

We expand the flow around a solution defined by a quasi-linear, yet
undefined polarization. In other terms we assume that there is little
ellipticity, i.e. $\theta_{0}\sim\pi/4$ and $d_{0}=0$. Besides,
we assume that the radius of the Stokes sphere is close to its steady
state value $S\sim S_{0}$ and consequently the carriers are also
around their equilibrium value $D=D_{0}$. In order to make apparent
the scale separation between the orientation of the polarization angle
and the rest of the variables, we introduce two time scales as
\begin{eqnarray}
\frac{d}{d\sigma} & = & \frac{\partial}{\partial\sigma_{0}}+\varepsilon\frac{\partial}{\partial\sigma_{1}}
\end{eqnarray}
as well as the following series expansion 
\begin{eqnarray}
S & = & S_{0}+\varepsilon S_{1}\left(\sigma_{0},\varepsilon\sigma_{1}\right)+\mathcal{O}\left(\varepsilon^{2}\right)\nonumber \\
\theta & = & \theta_{0}+\varepsilon\theta_{1}\left(\sigma_{0},\varepsilon\sigma_{1}\right)+\mathcal{O}\left(\varepsilon^{2}\right)\label{eq:sca}\\
D & = & D_{0}+\varepsilon D_{1}\left(\sigma_{0},\varepsilon\sigma_{1}\right)+\mathcal{O}\left(\varepsilon^{2}\right)\nonumber \\
d & = & d_{0}+\varepsilon d_{1}\left(\sigma_{0},\varepsilon\sigma_{1}\right)+\mathcal{O}\left(\varepsilon^{2}\right)\nonumber 
\end{eqnarray}
while $\psi_{\pm}\left(\sigma_{0},\varepsilon\sigma_{1}\right)$ is
not expanded perturbatively. As such, the orientation $\Phi=\psi_{+}-\psi_{-}$
can evolve freely between $0$ and $2\pi$. At the zeroth order,
we get the following system
\begin{equation}
\frac{\partial S_{0}}{\partial\sigma_{0}}=D_{0}S_{0}-\frac{\varepsilon_{g}P}{\omega_{r}}S_{0}^{2},\quad,\quad\frac{\partial D_{0}}{\partial\sigma_{0}}=1-S_{0},
\end{equation}
where we notice that with our scaling of parameters the relaxation
oscillations would not be damped if it wasn't for the non linear gain
saturation contribution. We find that $S_{0}=1$ and $D_{0}=\varepsilon_{g}P\omega_{r}^{-1}$.
Nicely the phases $\psi_{\pm}$ do not depend on the fast time scale
since the two zeroth order contributions cancel each other, i.e. 
\begin{eqnarray}
\frac{\partial\psi_{\pm}}{\partial\sigma_{0}} & = & 0
\end{eqnarray}
The first order problem reads on the fast time scale $\sigma_{0}$
\begin{eqnarray}
\frac{\partial S_{1}}{\partial\sigma_{0}} & \negthickspace\negthickspace=\negthickspace & -D_{0}S_{1}+D_{1}-2\frac{\gamma_{a}}{\omega_{r}}\cos\Phi-\frac{\varepsilon_{g}P}{2}D_{0}+\frac{\mathcal{F}_{0}}{\omega_{r}},\label{eq:S1}\\
\frac{\partial\theta_{1}}{\partial\sigma_{0}} & \negthickspace\negthickspace=\negthickspace & -\frac{d_{1}}{2}-\frac{\gamma_{a}}{\omega_{r}}\cos\Phi+\frac{\gamma_{p}}{\omega_{r}}\sin\Phi-\frac{\varepsilon_{g}P}{\omega_{r}}\theta_{1}+\frac{\mathcal{G}_{0}}{\omega_{r}},\label{eq:theta1}\\
\frac{\partial D_{1}}{\partial\sigma_{0}} & \negthickspace\negthickspace=\negthickspace & -\Gamma D_{0}-S_{1}+\frac{\varepsilon_{g}P}{2},\label{eq:D1}\\
\frac{\partial d_{1}}{\partial\sigma_{0}} & \negthickspace\negthickspace=\negthickspace & -\frac{\omega_{r}}{2}\frac{\gamma_{s}}{P}d_{1}+2\theta_{1}.\label{eq:d1}
\end{eqnarray}

We notice that Eqs.~(\ref{eq:S1},\ref{eq:D1}) and Eqs.~(\ref{eq:theta1},\ref{eq:d1})
correspond to two decoupled damped oscillators. While ellipticity
oscillations between $\left(d_{1},\theta_{1}\right)$ in Eqs.~(\ref{eq:theta1},\ref{eq:d1})
are strongly damped with a rate $\omega_{r}\gamma_{s}P$$^{-1}$,
the total intensity and carrier oscillations in Eqs.~(\ref{eq:S1},\ref{eq:D1})
need the damping due to the non linear saturation. This effect is
actually hidden in the fact that $D_{0}\neq0$. These two oscillators
are forced by the feedback terms $\mathcal{F}_{0}$ and $\mathcal{G}_{0}$,
the latter being evaluated at zeroth order since they are proportional
to $\eta$ and $\beta$ and therefore already first order quantities.
At this order in the expansion, these terms depend only the slow time
scale $\sigma_{1}$ through their dependence on variables $\psi_{\pm}$.
We can therefore readily solve Eqs.~(\ref{eq:S1}-\ref{eq:D1}) at
steady state and inject the adiabatic result in the first order problem
for $\psi_{\pm}$ on the slow time $\sigma_{1}$ that reads
\begin{eqnarray}
\frac{d\psi_{+}}{d\sigma_{0}} & = & \frac{\alpha}{2}\left(D_{1}+d_{1}\right)+\gamma_{a}\tan\theta\sin\Phi-\gamma_{p}\tan\theta\cos\Phi\nonumber \\
 & - & \alpha\frac{\varepsilon_{g}P}{4}D_{0}-\alpha\frac{D_{0}}{2}S_{1}+\alpha\frac{\varepsilon_{g}P}{\omega_{r}}\theta_{1}+\frac{F_{_{+}}^{0}}{\omega_{r}}\\
\frac{d\psi_{-}}{d\sigma_{0}} & = & \frac{\alpha}{2}\left(D_{1}-d_{1}\right)-\gamma_{a}\mathrm{cotan}\theta\sin\Phi-\gamma_{p}\mathrm{cotan}\theta\cos\Phi\nonumber \\
 & - & \alpha\frac{\varepsilon_{g}P}{4}D_{0}-\alpha\frac{D_{0}}{2}S_{1}-\alpha\frac{\varepsilon_{g}P}{\omega_{r}}\theta_{1}+\frac{F_{_{-}}^{0}}{\omega_{r}}
\end{eqnarray}

Upon replacing $D_{1}$ and $d_{1}$ from the steady state expression
of Eqs.~(\ref{eq:S1},\ref{eq:D1}), the several contributions due
to non linear saturation $\varepsilon_{g}$ cancel each other leaving
only the dichroism, the birefringence and the feedback terms to drive
the motion of $\psi_{\pm}$. After defining $u=\text{\ensuremath{\arctan\left(\alpha\right)}}$
and $\zeta=\arctan2\left(\gamma_{p},\gamma_{a}\right)$ we obtain
the phase model for $\psi_{+}$ and $\psi_{-}$ that reads, with $\left|z\right|=\sqrt{\gamma_{a}^{2}+\gamma_{p}^{2}}$
\begin{widetext}
\begin{eqnarray}
\frac{1}{\sqrt{1+\alpha^{2}}}\frac{d\psi_{\pm}}{dt} & = & \left|z\right|\sin\left(u\pm\psi_{+}\mp\psi_{-}-\zeta\right)+\frac{\beta}{2}\left[-\cos\left(\psi_{+}^{\tau_{r}}-\psi_{\pm}-a-u\right)+\cos\left(\psi_{-}^{\tau_{r}}-\psi_{\pm}-a-u\right)\right]\nonumber \\
 & \pm & \frac{\eta}{2}\left[\sin\left(\psi_{+}^{\tau_{f}}-\psi_{\pm}-\Omega-u\right)-\sin\left(\psi_{-}^{\tau_{f}}-\psi_{\pm}-\Omega-u\right)\right]\label{eq:psi_pm_final}
\end{eqnarray}

\end{widetext}where we reintroduced the original time scale in Eq.~(\ref{eq:psi_pm_final})
to clarify that the characteristic time scale is governed by the amplitude
of the terms $\gamma_{a},\gamma_{p},\eta$ and $\beta$. Notice however
that the dynamics may not be restricted to a particularly slow time
scales as the main requirement for our analysis to hold is for the
four aforementioned parameters to be smaller than the damping of the
relaxation oscillations which is typically $10\sim15\,$GHz, hence
studying multi-GHz dynamics would still possible within our simplified
approach. At last, we stress that it is not because we reduced the
dynamics to the evolution of the equatorial component on the Stokes
Sphere $\Phi$ that the dynamics is confined to purely linear polarization.
Indeed, we can express $\theta$ and $n$ solving Eqs.~(\ref{eq:theta1},\ref{eq:d1})
as
\begin{equation}
\theta\left(\psi_{+}-\psi_{-}\right)=\frac{\pi}{4}+\frac{\mathcal{G}_{0}-\left|z\right|\cos\left(\zeta+\psi_{+}-\psi_{-}\right)}{P\left(2\gamma_{s}^{-1}+\varepsilon_{g}\right)}\:,\label{eq:slave_ellipticity}
\end{equation}
and similarly for the spin imbalance $n=2P\left(\theta-\frac{\pi}{4}\right)\gamma_{s}^{-1}$.
Equivalent yet cumbersome expressions for the intensity and total
carrier variations can be obtained in the same way. From Eq.~(\ref{eq:slave_ellipticity}),
it is apparent that the typical deviations of $\theta$ with respect
to $\pi/4$ and the spin-imbalance $\delta n$ are 
\begin{equation}
\delta\theta\sim\pm\pi/10\;,\; n\sim\pm0.1.
\end{equation}
Notice that $\delta\theta=\pi/4$ would correspond to a purely circular
emission state.

\section{Results}

The modal structure of the VCSEL submitted to optical feedback and
cross-injection is more conveniently studied by defining the half
sum $\Sigma=\left(\psi_{+}+\psi_{-}\right)/2$ and the difference
$\Phi=\psi_{+}-\psi_{-}$. In the case of a mono-mode solution the
difference $\Phi$ fixes the orientation of the quasi-linear polarization
and reaches a fix point while the half sum drifts at the frequency
of the mode under consideration. After some trigonometric simplifications
Eq.~(\ref{eq:psi_pm_final}) transforms into \begin{widetext} 
\begin{eqnarray}
\frac{1}{\sqrt{\cdot}}\frac{d\Sigma}{ds} & \negthickspace=\negthickspace & \left|z\right|\cos\Phi\sin\left(u-\zeta\right)-\eta\sin\frac{\Phi^{\tau_{f}}}{2}\sin\frac{\Phi}{2}\sin\left(u+\Omega+\Sigma-\Sigma^{\tau_{f}}\right)-\beta\cos\frac{\Phi}{2}\sin\frac{\Phi^{\tau_{r}}}{2}\sin\left(u+a+\Sigma-\Sigma^{\tau_{r}}\right),\label{eq:Sig}\\
\frac{1}{2\sqrt{\cdot}}\frac{d\Phi}{ds} & \negthickspace=\negthickspace & \left|z\right|\sin\Phi\cos\left(u-\zeta\right)+\eta\sin\frac{\Phi^{\tau_{f}}}{2}\cos\frac{\Phi}{2}\cos\left(u+\Omega+\Sigma-\Sigma^{\tau_{f}}\right)-\beta\sin\frac{\Phi}{2}\sin\frac{\Phi^{\tau_{r}}}{2}\cos\left(u+a+\Sigma-\Sigma^{\tau_{r}}\right),\label{eq:Phi}
\end{eqnarray}
\end{widetext} with $\sqrt{\cdot}=\sqrt{1+\alpha^{2}}$. Interestingly,
the symmetry properties of Eqs.~(\ref{eq:Sig},\ref{eq:Phi}) are
not equivalent with respect to $\Sigma$ and $\Phi$. While Eqs.~(\ref{eq:Sig},\ref{eq:Phi})
are phase invariant with respect to $\Sigma$, this is not the case
for $\Phi$. It is an expected result since these two phases do not
have the same physical meaning, $\Sigma$ is an optical phase while
$\Phi$ is an orientation angle.

Mono-modes solutions correspond to $\Sigma=\omega t$ while $\Phi$
is a constant. Besides the value of $\omega$ and $\Phi$ we will
represent the associated ellipticity $\theta$ using Eq.~(\ref{eq:slave_ellipticity}),
exploiting that the expression of $\mathcal{G}_{0}$ in the case of
a monochromatic solution simply read
\begin{eqnarray}
\mathcal{G}_{0} & \negthickspace\negthickspace=\negthickspace\negthickspace & 2\beta\sin^{2}\frac{\Phi}{2}\sin\left(\omega\tau_{r}+a\right)-\eta\sin\Phi\sin\left(\omega\tau_{f}+\Omega\right)
\end{eqnarray}

\subsection{Particular cases}

Before studying the general case, we will recover several known situations
as particular cases in the absence of any feedback or with only PSF.

\paragraph{Solitary VCSEL :}

We first discuss the stability of the solitary VCSEL that is governed
by a single equation for $\Phi$ Eq.~(\ref{eq:Phi}) that reads after
simplification 
\begin{eqnarray}
\frac{d\Phi}{ds} & = & 2\left(\gamma_{a}+\alpha\gamma_{p}\right)\sin\Phi,\label{eq:Phi_Simple}
\end{eqnarray}
the solutions $\Phi_{x}=0$ and $\Phi_{y}=\pi$ correspond respectively
to a saddle and a node when $\gamma_{a}+\alpha\gamma_{p}>0$, and
vice-versa in the opposite case. The frequency of such two modes can
be deduced from Eq.~(\ref{eq:Sig}) that reads 
\begin{eqnarray}
\frac{d\Sigma}{ds} & = & \left(\alpha\gamma_{a}-\gamma_{p}\right)\cos\Phi_{x,y}=\omega_{x,y}.\label{eq:Sig_Simple}
\end{eqnarray}

\begin{figure}[tp]
\begin{centering}
\includegraphics[bb=0bp 0bp 243bp 140bp,width=0.49\textwidth]{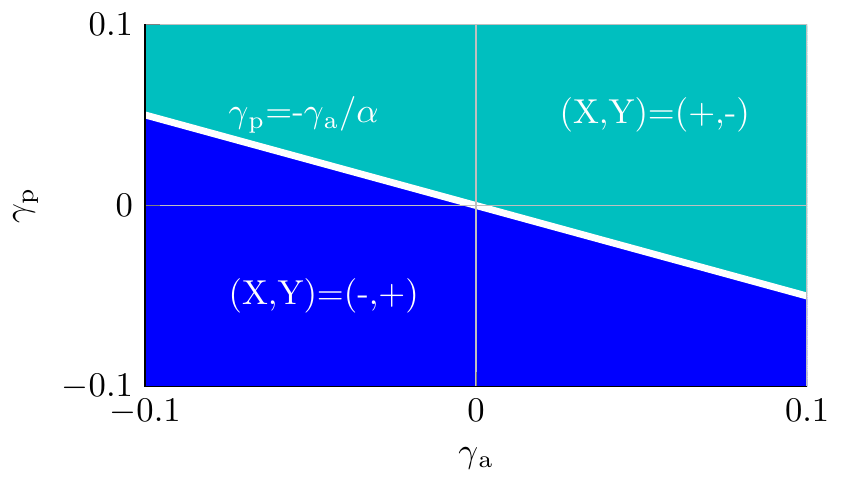} 
\par\end{centering}

\centering{}\caption{(Color Online): Stability diagram of the solitary VCSEL far from threshold.
The condition $\gamma_{a}+\alpha\gamma_{p}=0$ separates the two parameter
regions where the stable emission is along the Y or the X axis. Stable
(resp. instable ) emission is depicted by a $-$ sign (resp. a $+$
sign). \label{fig:theo_0b} }
\end{figure}

The stability diagram is depicted in Fig.~\ref{fig:theo_0b}. Notice
that the stability diagram of Fig.~\ref{fig:theo_0b} is much simpler
than for instance the analysis performed in \cite{EDP-PRA-99}. For
instance there is no bistability in our case like since we are far
from threshold. However, we checked numerically the accuracy of the
stability information predicted by Eq.~\ref{eq:Phi_Simple} and found
a good agreement, in the sense that the condition $\gamma_{a}+\alpha\gamma_{p}=0$
indeeds separate mono-stable emission along LP-x from LP-y. However,
even far from threshold, a very small region of bistability was found
in the vicinity of the line $\gamma_{a}+\alpha\gamma_{p}=0$ represented
in Fig.~\ref{fig:theo_0b}. Since our simplified model is build on
a perturbative expansion, it is not abnormal that some small, usually
negligible terms dominate the stability close to the special conditions
in parameter space for which the first order term vanishes.

\paragraph{LP-x emission :}

Irrespectively of the values of $\eta$ and $\beta$, pure emission
along the $X-$axis of the solitary VCSEL is always possible since
the existence of this mode is obviously not affected by optical feedback
into $Y$ and cross-injection of $Y$ into $X$. This formally corresponds
to the solution $\Phi=0$ which solves Eq.~(\ref{eq:Phi}) while
Eq.~(\ref{eq:Sig}) reduces to the expression of the frequency of
the $X$ solution at frequency $-\gamma_{p}$ pulled by the interplay
of the dichroism $\gamma_{a}$ and $\alpha$, i.e. 
\begin{eqnarray}
\omega_{x} & = & \sqrt{1+\alpha^{2}}\left|z\right|\sin\left(u-\zeta\right)=\alpha\gamma_{a}-\gamma_{p}
\end{eqnarray}
where we used trigonometrical identifies to simplify the last result.

\paragraph{LP-y Lang-Kobayashi modes :}

Secondly, on the case where there is only optical feedback, i.e. $\beta=0$,
the solution $\Phi=\pi$ solves Eq.~(\ref{eq:Phi}). This case corresponds
to a linear polarization along the $Y$ axis and Eq.~(\ref{eq:Sig})
reduces exactly to the locus of the modes of the Lang-Kobayashi model
\begin{eqnarray}
\omega_{y}-\gamma_{p}+\alpha\gamma_{a}+\eta\sqrt{1+\alpha^{2}}\sin\left(u+\Omega+\omega_{y}\tau_{f}\right) & \negthickspace\negthickspace=\negthickspace\negthickspace & 0,
\end{eqnarray}
at the only difference that the ellipse of the modes is shifted by
the birefringence $\gamma_{p}$ as well as the contribution $\alpha\gamma_{a}$.

\begin{figure}[tp]
\begin{centering}
\includegraphics[width=0.49\textwidth]{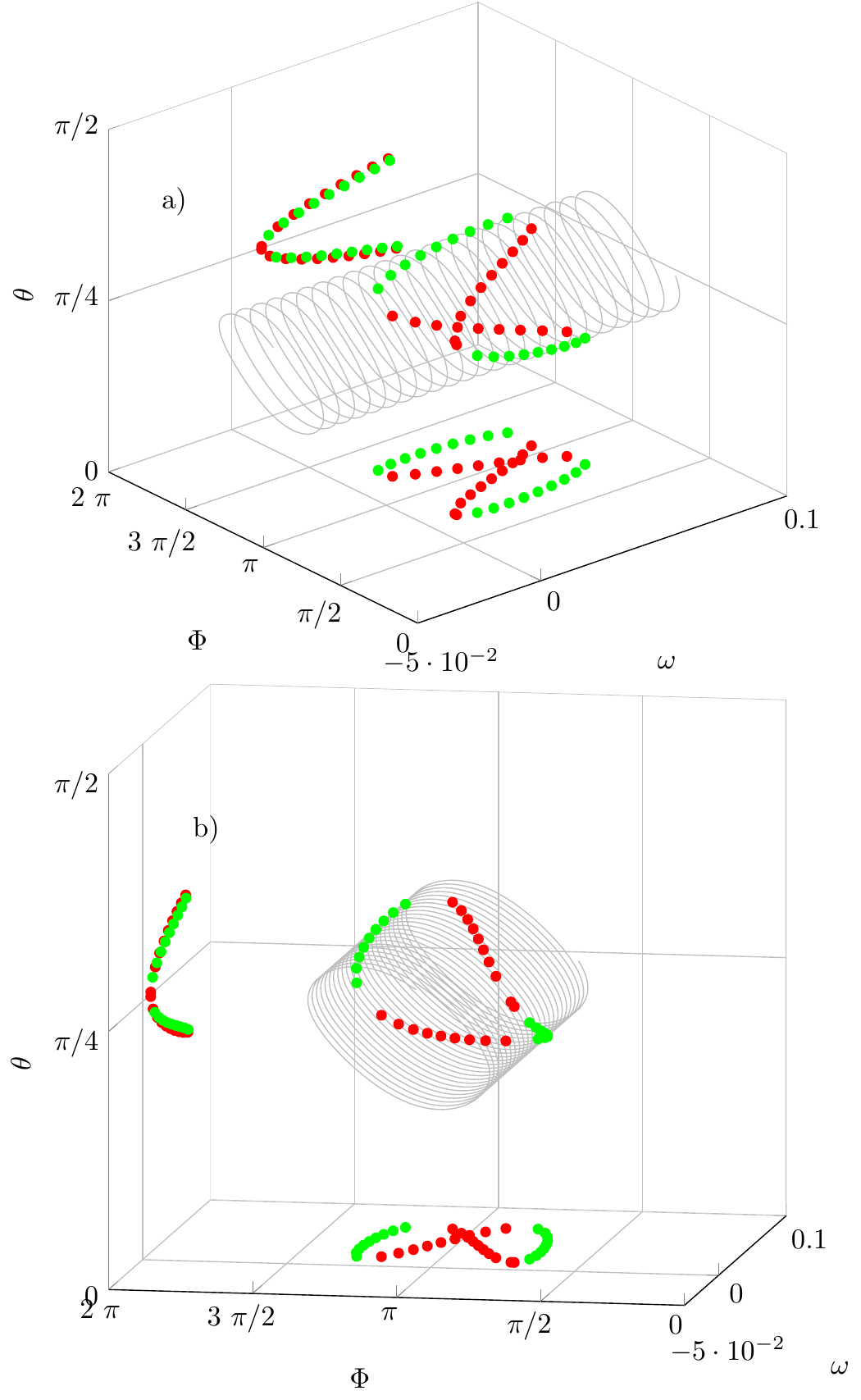} 
\par\end{centering}

\centering{}\caption{(Color Online): Monochromatic solutions of Eqs.~(\ref{eq:Sig},\ref{eq:Phi})
for $\eta=0$, $\beta=0.05$ and $\tau_{r}=1000$. The solutions are
arranged around a tube defined by the function $\Phi\left(\omega\right)$
and $\theta\left(\omega\right)$ assuming $\omega$ a continuous variable.
Stable and unstable solutions are represented in green and red, respectively.
The number of stable and unstable solutions $\left(S,U\right)$ is
$\left(18,19\right)$ \label{fig:theo_1} }
\end{figure}

In the general case, the presence of cross-polarization make it so
that the orientation cannot perfectly align with the $Y-$axis, hence
$\Phi\neq0,\pi$. This allows simplifying Eq.~(\ref{eq:Phi}) dividing
by $\sin\Phi$ and to express the orientation as a function of the
frequency as 
\begin{eqnarray}
\frac{\Phi}{2} & \negthickspace\negthickspace=\negthickspace\negthickspace & \arctan\frac{2\left|z\right|\cos\left(u-\zeta\right)+\eta\cos\left(u+\Omega+\omega\tau_{f}\right)}{\beta\cos\left(u+a+\omega\tau_{r}\right)}.
\end{eqnarray}
Such value of $\Phi$ must be replaced in Eq.~(\ref{eq:Sig}) to
yield the locus of the quasi-linear modes as solutions of

\begin{eqnarray}
\frac{\omega}{\sqrt{1+\alpha^{2}}} & = & \left|z\right|\cos\left[\Phi\left(\omega\right)\right]\sin\left(u-\zeta\right)\nonumber \\
 & - & \eta\sin^{2}\left[\frac{\Phi\left(\omega\right)}{2}\right]\sin\left(u+\Omega+\omega\tau_{f}\right)\nonumber \\
 & - & \beta\sin\left[\Phi\left(\omega\right)\right]\sin\left(u+a+\omega\tau_{r}\right)\label{eq:locus}
\end{eqnarray}

\subsection{Cross polarization only}

In the case where the VCSEL is submitted to only cross-polarization
re-injection, i.e. $\eta=0$, we recover the results of \cite{MGJ-PRA-07}
as a particular case. In was shown in \cite{MGJ-PRA-07} that modes
appears as saddle-node bifurcations for increasing values of $\beta$,
in addition to the pure $Y$ solution of the solitary VCSEL that evolves
due to the influence of $\beta$. It entails that there is usually
$N/2+1$ modes and $N/2$ anti-modes. Notice however that in \cite{MGJ-PRA-07}
the modal structure was found by solving exactly the Eqs.~(\ref{eq:Epm},\ref{eq:Dpm})
and without exploiting the particular scaling of the parameters. As
such the solution was found as a determinant of a complex system of
equations. Here, one is able to recognizes by inspecting Eq.~(\ref{eq:locus})
that the locus for the modes is in essence very similar to the one
of the Lang-Koyabashi model, i.e. a transcendental equation defining
the frequencies. 

In addition to the solution defined by the triplet $\left(\omega,\Phi,\theta\right)$
we represented $\Phi\left(\omega\right)$ and $\theta\left(\omega\right)$
as continuous functions of $\omega$, the reason for doing so will
be clarified in the next section. This corresponds to the black lines
in Fig.~\ref{fig:theo_1}. The information regarding the stability
of the solutions was found using DdeBiftool \cite{DDEBT} on the Eqs.~(\ref{eq:Sig},\ref{eq:Phi}).
Importantly, one notices in Fig.~\ref{fig:theo_1} that purely linear
emission corresponding to solutions for which with $\theta=\pi/4$
is possible. This demonstrates that cross polarization can have the
effect to simply rotate the direction of emission. The modes ``ellipse''
is centered around the frequency $\omega_{y}=\gamma_{p}-\alpha\gamma_{a}$
of the pure $Y$ emission state and the states closer to $\Phi=\pi$,
i.e. whose polarization is close to the $Y$ axis, are the one the
most unstable. In Fig.~\ref{fig:theo_1}, the external part of the
projection in the $\left(\omega,\Phi\right)$ plane (the stable modes
in green) corresponds to the orientations farther from $Y$. Such
modes can also present a small ellipticity. Notice that in Fig.~8
of \cite{MGJ-PRA-07}, the variable $n$, or equivalently here $\theta$,
presented a height shaped curve. This was due to the larger values
of $\beta=0.25$ used in \cite{MGJ-PRA-07}. 

\begin{figure}[tp]
\begin{centering}
\includegraphics[width=0.49\textwidth]{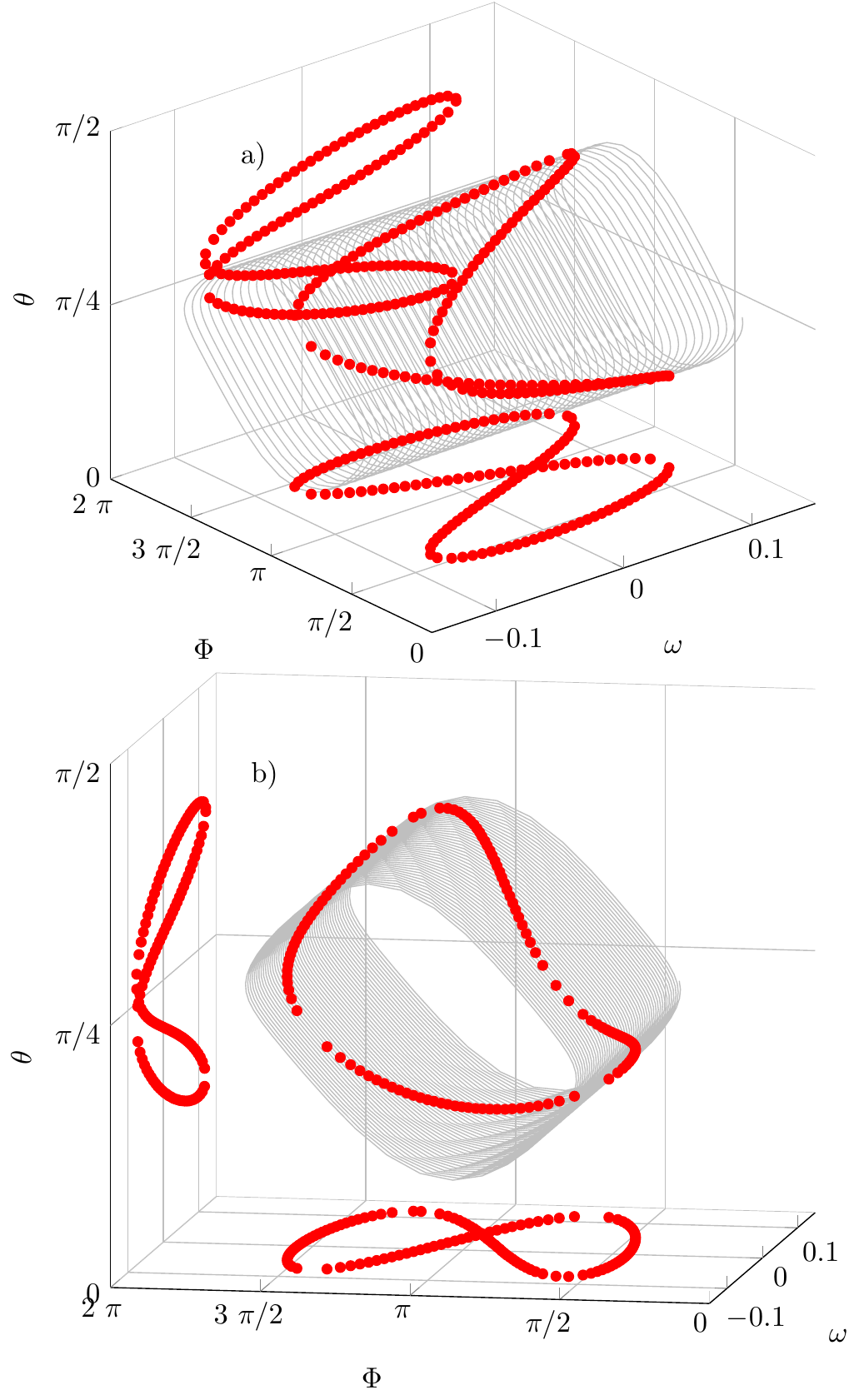} 
\par\end{centering}

\centering{}\caption{(Color Online): Monochromatic solutions of Eqs.~(\ref{eq:Sig},\ref{eq:Phi})
for $\eta=0$, $\beta=\beta^{\star}+0.01$ and $\tau_{r}=1000$. There
are 127 unstable solutions. Stable and unstable solutions are represented
in green and red, respectively. \label{fig:theo_2} }
\end{figure}

\begin{figure}[tp]
\begin{centering}
\includegraphics[clip,width=0.49\textwidth]{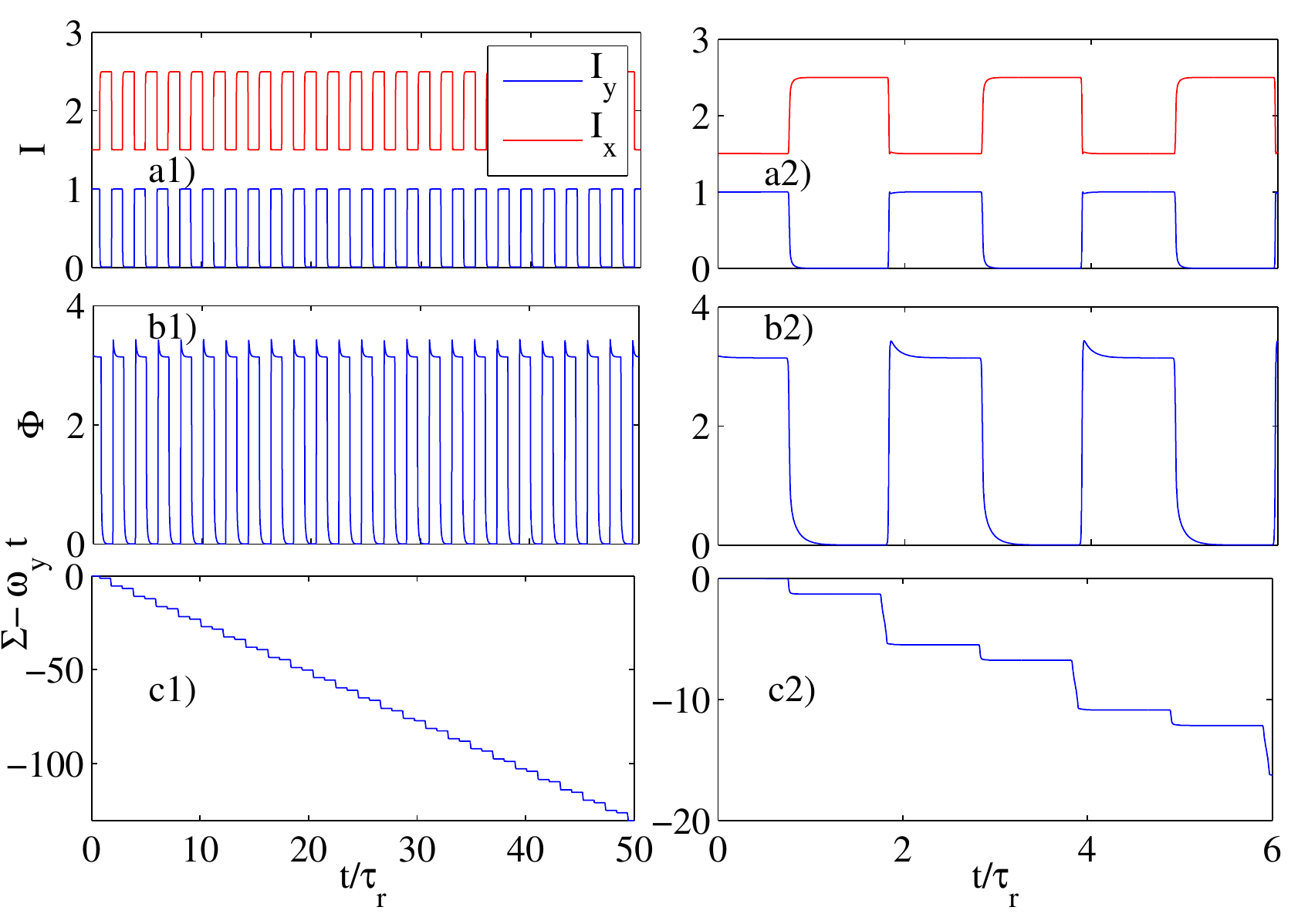} 
\par\end{centering}

\centering{}\caption{(Color Online): Square-wave switching as phase kinks. The panels a1)
and a2) describe the intensity of the X and Y polarization over different
time scales reconstructed from the polarization orientation $\Phi.$
The traces are shifted for clarity, and are actually in perfect anti-phase.
The panels b1) and b2) represent the polarization orientation $\Phi$
while panels c1) and c2) describe the optical phase $\Sigma$ to which
we subtracted a drift $\omega_{y}t$. The parameters are $\eta=0$,
$\beta=\beta^{\star}+4\times10^{-3}$ and $\tau_{r}=1000$. \label{fig:theo_3} }
\end{figure}

For increasing values of $\beta$ more modes are created via saddle-node
bifurcations and the ellipse grows, yet more and more of the stable
modes at the exterior of the eight shaped projection becomes unstable
up to the critical value $\beta^{\star}=2\sqrt{\gamma_{a}^{2}+\gamma_{p}^{2}}$
where all the modes are unstable, see Fig.~\ref{fig:theo_1}. This
corresponds to the onset of square wave switching \cite{GES-OL-06,MGJ-PRA-07}
where the two orthogonal polarizations $X$ and $Y$ alternate a cycle
of on-off emission in anti-phase and at a period close to twice the
delay imposed by cross-polarization re-injection $\tau_{r}$. Such
dynamics at twice the delay is depicted in Fig.~\ref{fig:theo_3}.

Here we reconstructed the intensity of the X and Y components via
the formula $I_{x}\sim\left|1+e^{i\Phi}\right|^{2}$ and $I_{y}\sim\left|1-e^{i\Phi}\right|^{2}$.
By construction, the anti-phase between the two polarization is perfect
as one can notice in Fig.~\ref{fig:theo_3}a). For $\beta>\beta^{\star}$
we found in Fig.~\ref{fig:theo_3}a) a strongly non linear limit
cycle composed of two plateaus whose period is close to twice the
re-injection delay $\tau_{r}$ in agreement with the results of \cite{GES-OL-06,MGJ-PRA-07,MJG-PRA-13}.
Interestingly, we notice that it is possible to re-interpret such
anti-phase dynamics for the intensities as a pure phase dynamics.
Indeed we describe in Fig.~\ref{fig:theo_3}b) the temporal evolution
of $\Phi$ which consists in kinks between $\pi$ and $0$ and $\pi$
again. The evolution of the global phase $\Sigma$ consists mainly
in a drift at a frequency $\omega_{y}$, i.e. the frequency of the
solitary VCSEL on the Y mode. This is consistent with the fact that
the first plateau corresponds to pure $Y$ emission while the second
one consists in the $Y$ mode performing injection locking into the
$X$ polarization. In both cases the frequency of emission is $\omega_{y}$.
Interestingly, once this drift is removed, a residual phase kink can
be observed in $\Sigma$. However, such residual kinks in $\Sigma$
are slaved to the square-wave switching and actually can be decoupled
from the dynamics. For $\beta\gg\beta^{\star}$ there is an excellent
agreement between the full solution of Eqs.~(\ref{eq:Sig},\ref{eq:Phi})
and the one of Eq.~(\ref{eq:Phi}) in which we performed the substitution
$\Sigma-\Sigma^{\tau_{r}}\rightarrow\omega_{y}\tau_{r}$ demonstrating
that the square wave phenomena can be reduced to a single equation
with delay of the type
\begin{eqnarray}
\frac{d\Phi}{dt'} & = & \sin\frac{\Phi}{2}\left(\cos\frac{\Phi}{2}-A\sin\frac{\Phi^{\tau'_{r}}}{2}\right)\label{eq:Phi-simple}
\end{eqnarray}
where we scale time for clarity and defined an effective parameter
\begin{eqnarray}
A & \negthickspace=\negthickspace & \frac{\beta}{2}\frac{\sqrt{1+\alpha^{2}}}{\gamma_{a}+\alpha\gamma_{p}}\cos\left[u+a+\left(\gamma_{p}-\alpha\gamma_{a}\right)\tau_{r}\right]
\end{eqnarray}
Inspection of Eq.~(\ref{eq:Phi-simple}) reveals that solutions composed
of plateaus of duration $\tau_{r}$ for which either $\Phi=0$ or
$\Phi=\pi$ are indeed possible.

\begin{figure*}[!]
\begin{centering}
\includegraphics[bb=0bp 0bp 286bp 279bp,width=0.49\textwidth]{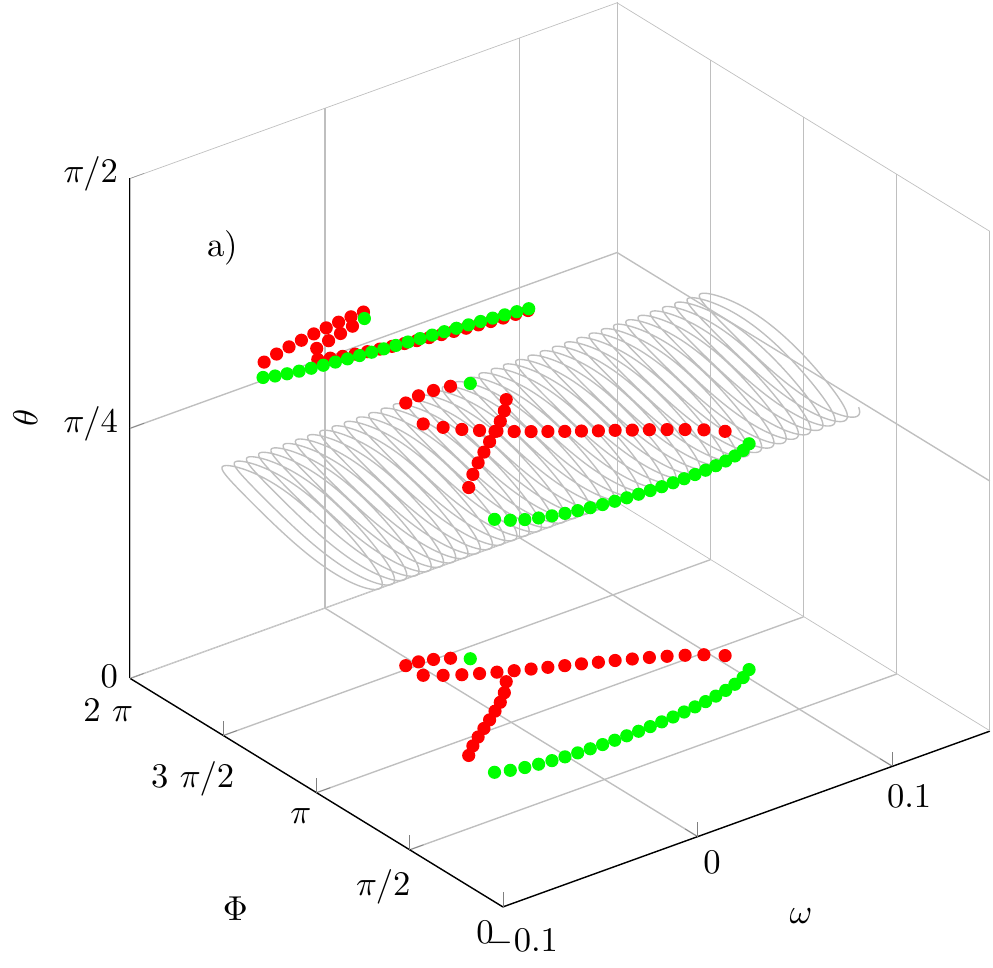}\includegraphics[bb=0bp 0bp 286bp 279bp,width=0.5\textwidth]{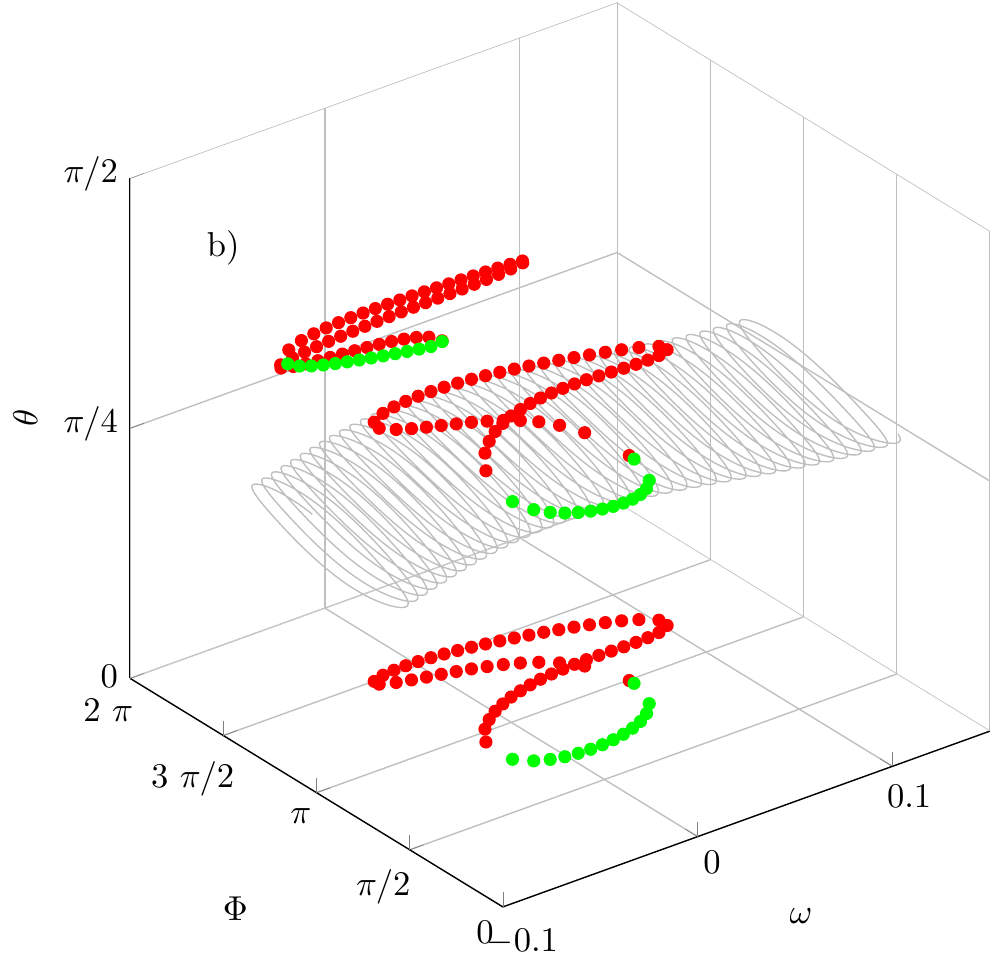} 
\par\end{centering}

\vspace{1.5cm}

\begin{centering}
\includegraphics[bb=0bp 0bp 286bp 279bp,width=0.49\textwidth]{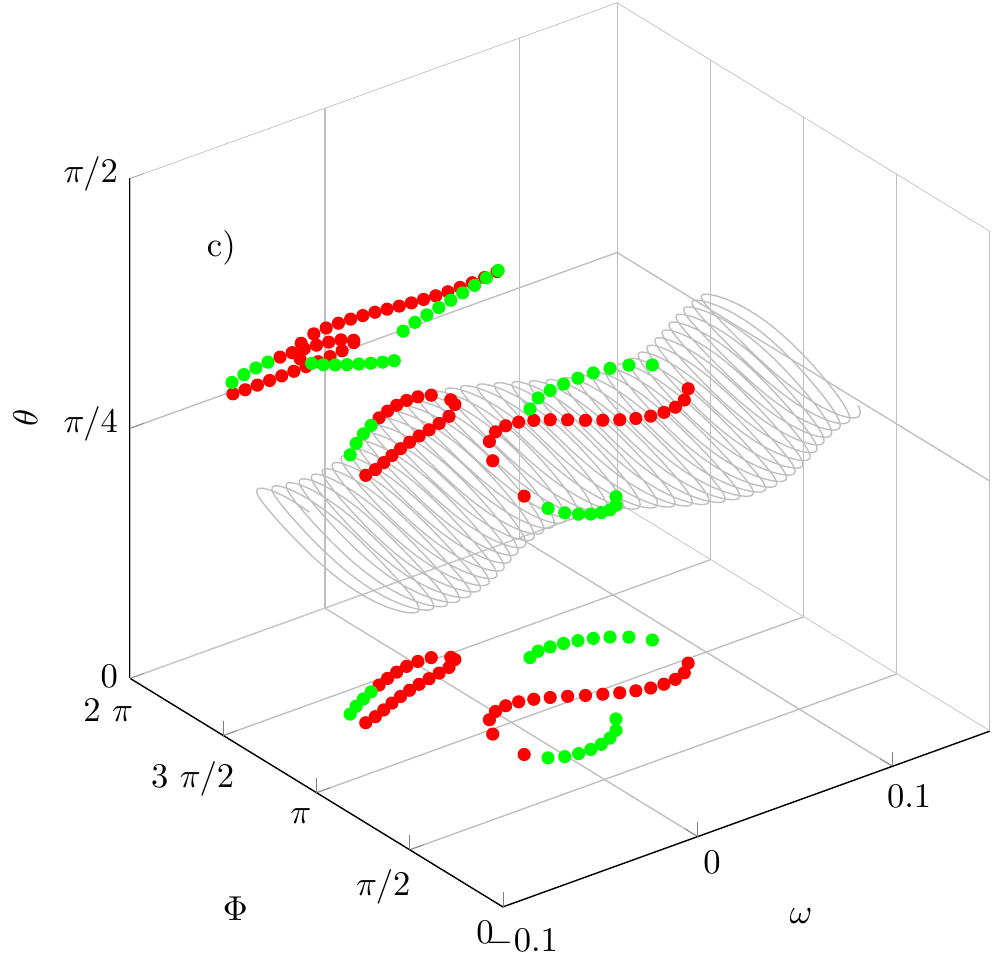}
\includegraphics[bb=0bp 0bp 286bp 279bp,width=0.49\textwidth]{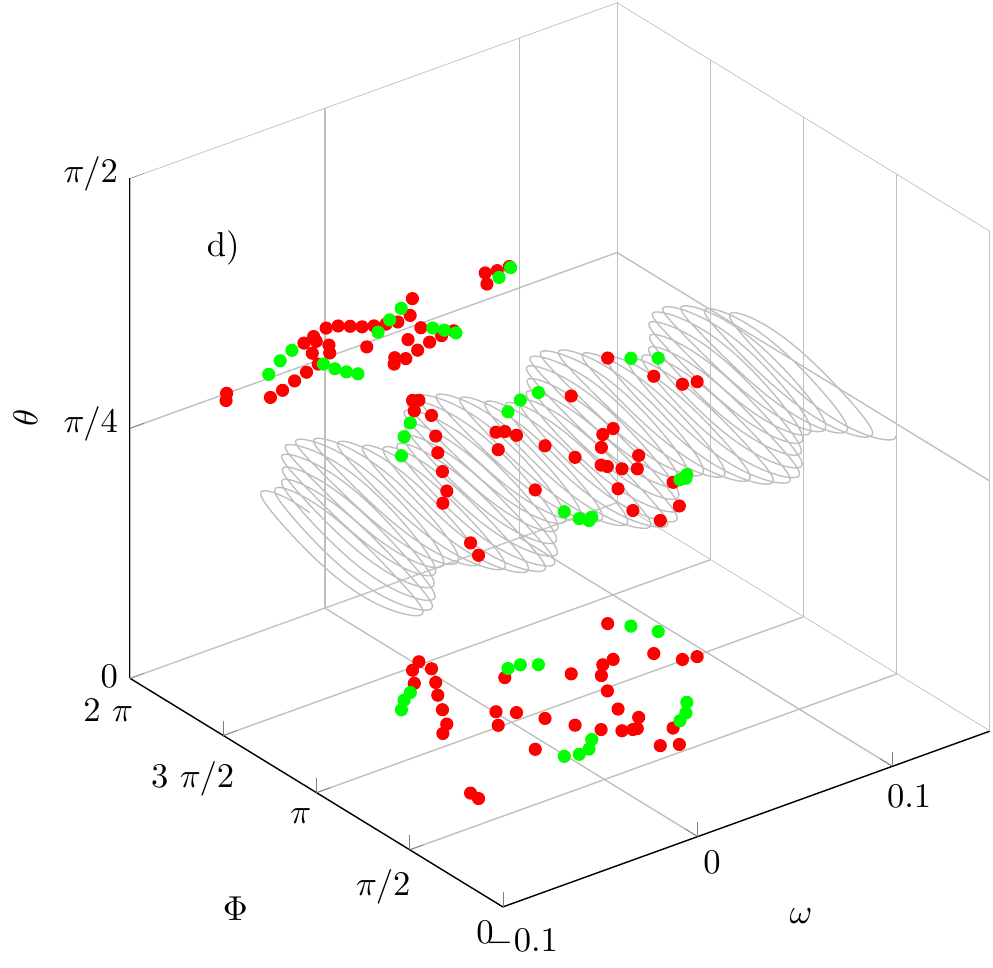} 
\par\end{centering}

\centering{}\caption{(Color Online): Monochromatic solutions of Eqs.~(\ref{eq:Sig},\ref{eq:Phi})
for $\eta=0.025$, $\beta=0.05$ and $\tau_{r}=1000$. Panels a),
b), c) and d) correspond to $\Delta\tau=\tau_{f}-\tau_{r}=0$, $\Delta\tau=20$,
$\Delta\tau=40$, and $\Delta\tau=100$ for which the number of stable
and unstable solutions $\left(S,U\right)$ is $\left(24,31\right)$,
$\left(14,55\right)$, $\left(21,36\right)$ and $\left(15,36\right)$
\label{fig:theo_4} }
\end{figure*}

\begin{figure*}[!]
\begin{centering}
\includegraphics[bb=0bp 0bp 286bp 279bp,width=0.49\textwidth]{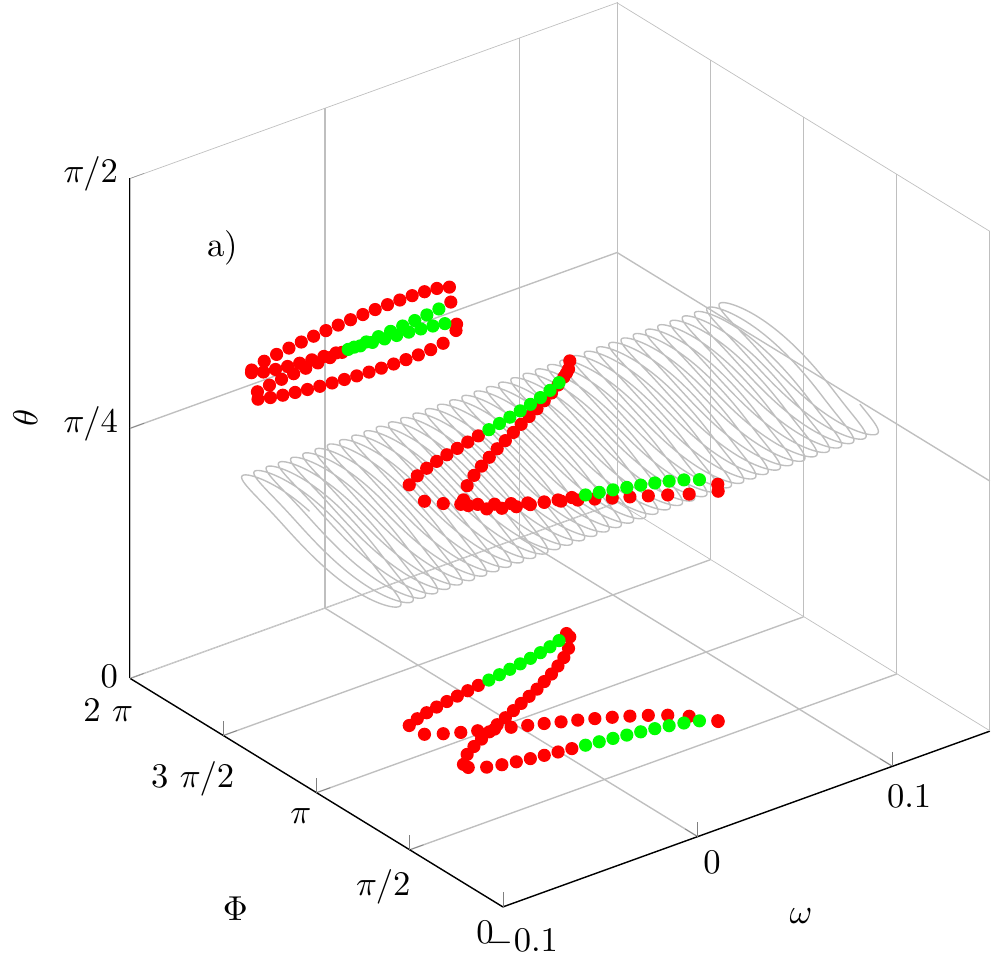}\includegraphics[bb=0bp 0bp 286bp 279bp,width=0.49\textwidth]{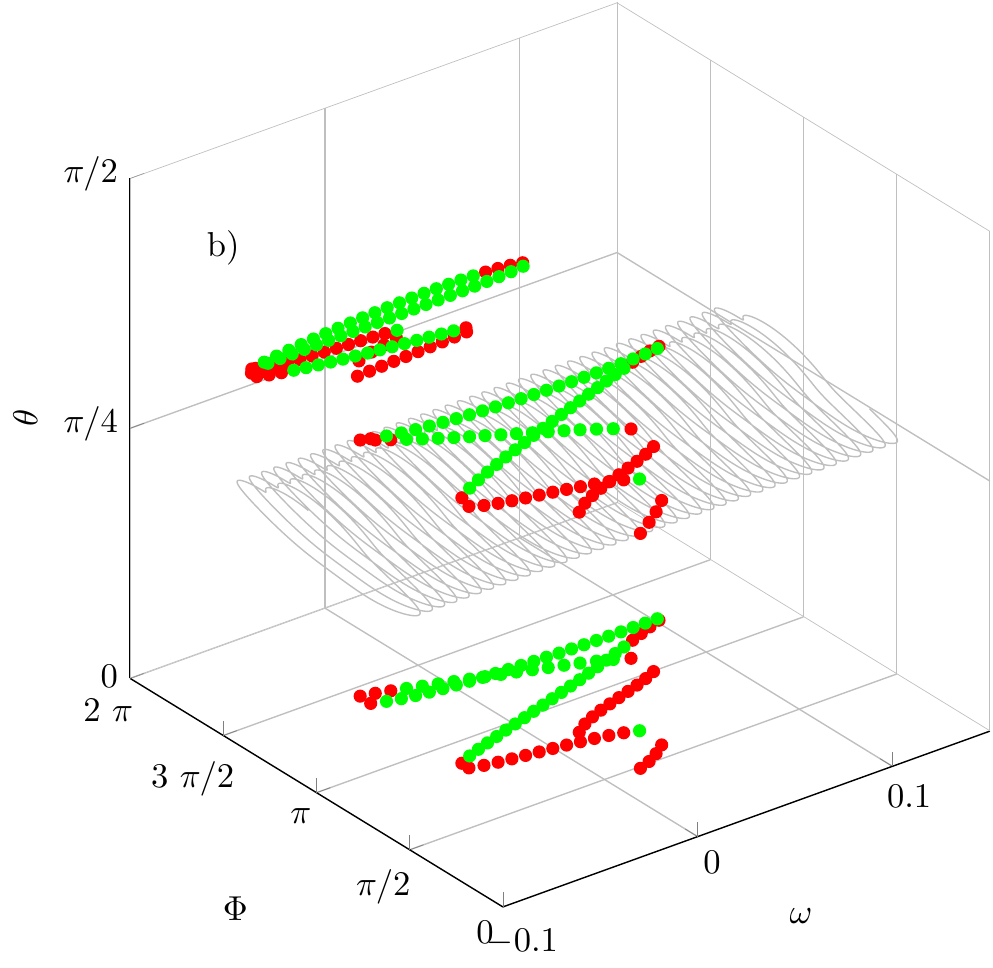} 
\par\end{centering}

\vspace{1.5cm}

\begin{centering}
\includegraphics[bb=0bp 0bp 286bp 279bp,width=0.49\textwidth]{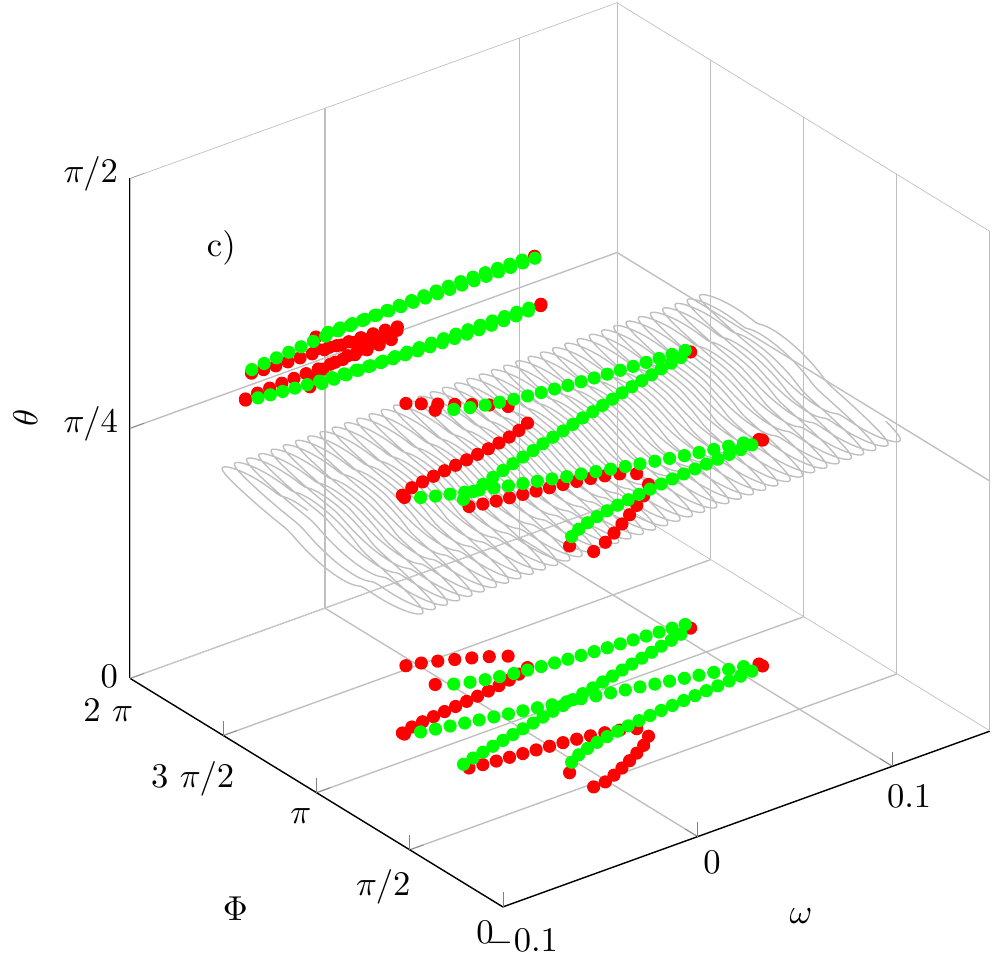}
\includegraphics[bb=0bp 0bp 286bp 279bp,width=0.49\textwidth]{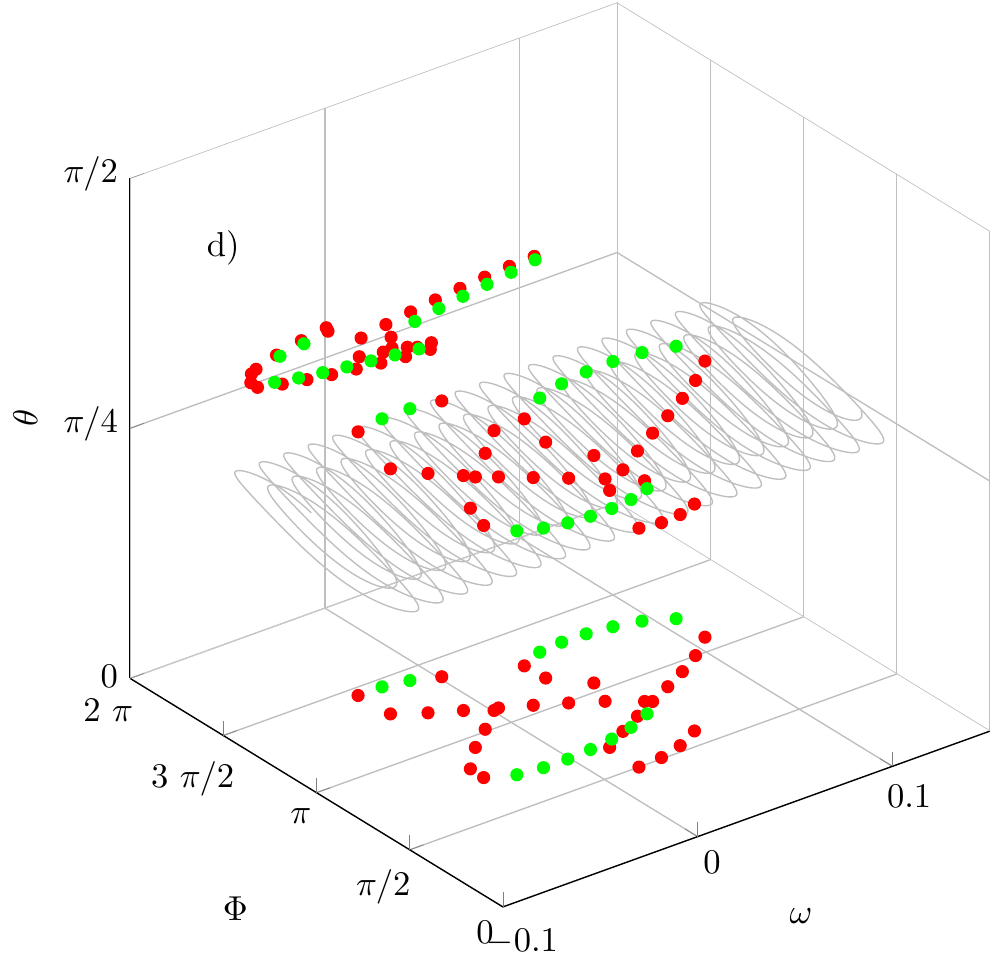} 
\par\end{centering}

\centering{}\caption{(Color Online): Monochromatic solutions of Eqs.~(\ref{eq:Sig},\ref{eq:Phi})
for $\eta=0.025$, $\beta=0.05$ and $\tau_{r}=1000$. Panels a),
b), c) and d) correspond to $\tau_{f}=2\tau_{r}$, $\tau_{f}=3\tau$$_{r}$,
$\tau_{f}=4\tau_{r}$, and $\tau_{f}=\tau_{r}/2$ for which the number
of stable and unstable solutions $\left(S,U\right)$ is $\left(17,52\right)$,
$\left(55,36\right)$, $\left(82,47\right)$ and $\left(15,30\right)$
\label{fig:theo_5} }
\end{figure*}

\subsection{Cross polarization and feedback}

In the general case where both $\eta$ and $\beta$ are non zero,
the modal structure depends critically on the difference between the
two delays. For small deviations for the situation $\tau_{f}=\tau_{r}$
the height shaped mode ellipse distorts and break in several parts
as depicted in Fig.~\ref{fig:theo_4}. Here, one notice that the
tubular structure that supports the mode acquires a modulation that
is proportional to the difference between the two delays. For small
differences between the two delays like in Fig.~\ref{fig:theo_4}a)
and Fig.~\ref{fig:theo_4}b) the mode ``ellipse'' is deformed.
For larger differences like e.g. in Fig.~\ref{fig:theo_3}c) the
modes position break into several sub families. Increasing the difference
between the two delays beyond Fig.~\ref{fig:theo_4}d) would give
a modal structure that would seems like random points (not shown)
if observed only through a projection in the $\left(\omega,\Phi\right)$
plane. We depicted in Fig.~\ref{fig:theo_4} a sequence for which
$\tau_{f}$ is increased above $\tau_{r}$, yet a similar scenario
is found for $\tau_{f}<\tau_{r}$. 

Once it is understood that the tubular structure oscillate at a frequency
given by the difference between the two delays, one may foresee the
existence of ``revivals'' of relatively simple modal structures
for specific ratio between the two delays. Indeed we show in Fig.~\ref{fig:theo_5}
that a regular structure exists whenever the feedback delay is an
integer of the cross polarization delay. Similarly, some simple structure
were also found when $\tau_{f}=\tau_{r}/n$ and we depicted in Fig.~\ref{fig:theo_5}d)
the case $\tau_{f}=\tau_{r}/2$. 

\newpage{}

\subsection{Influence of optical feedback onto the square wave switching}

The second plateau of the square wave dynamics described in \cite{MGJ-PRA-07,MJG-PRA-13}
consists in the strong mode (say LP-y) injection locking the weak
mode (say LP-x). At the end of the second plateau, the transitory
dynamics can be understood as an escape of the vicinity of a possibly
weakly repulsive saddle. During such escape the system is very sensible
to noise which induces strong period jitter in the square wave signal.
The non vanishing time needed to perform such an escape explains why
the period of the square oscillation is always slightly superior to
twice the XPR delay $\tau_{r}$. 

\begin{figure}[tp]
\begin{centering}
\includegraphics[bb=0bp 0bp 552bp 333bp,width=0.49\textwidth]{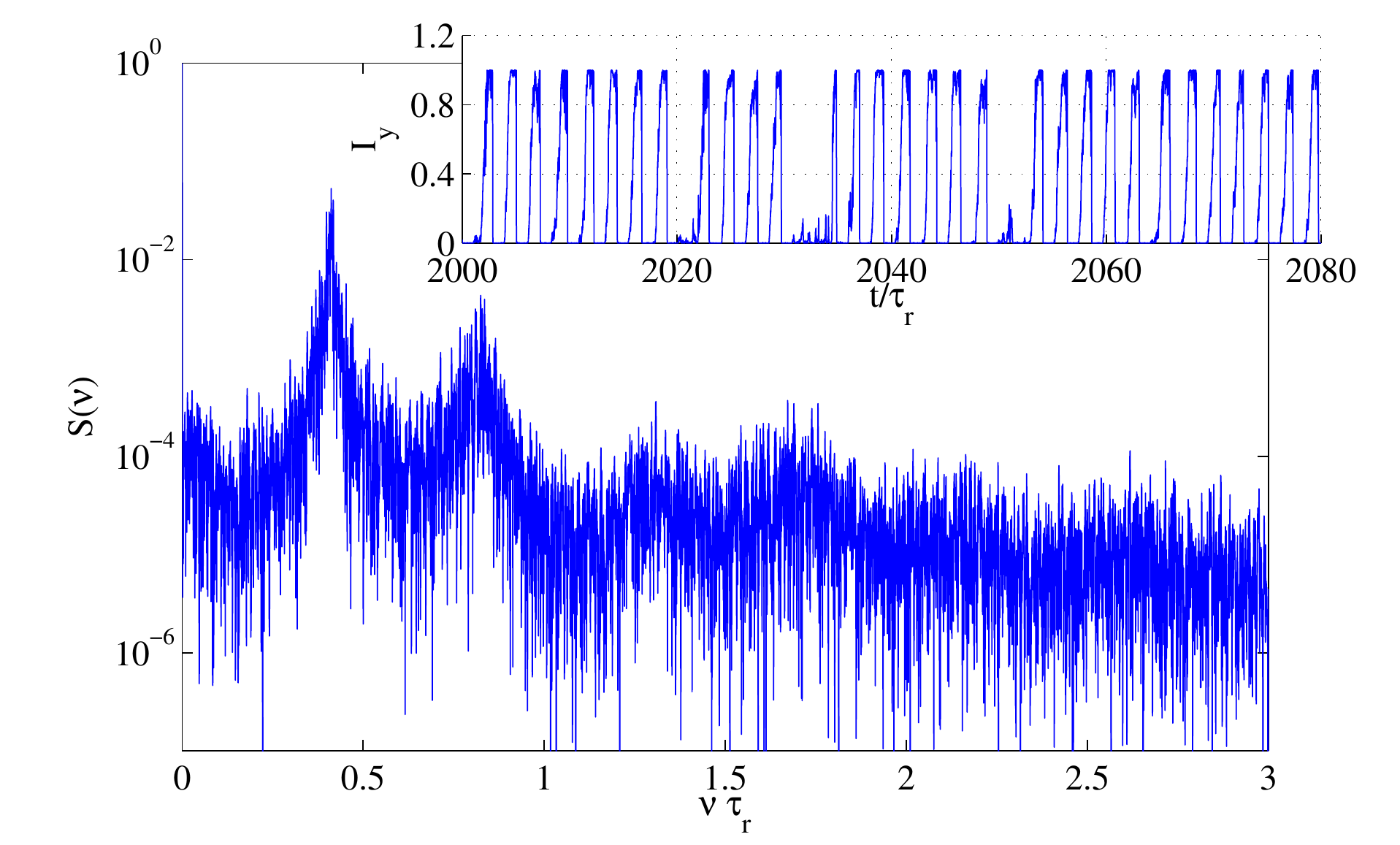}
\par\end{centering}

\centering{}\caption{(Color Online): Square wave switching dynamics (inset) and power spectrum
for $\eta=0$, $\gamma_{a}=-0.09$,$\beta=0.25$ and $\tau_{r}=500$
and power spectrum. The signal is very irregular and the period is
noticeably slower than $2\tau_{r}$. The power spectrum does not show
the signature of a square wave signal with only odd harmonics. \label{fig:theo_6a} }
\end{figure}

The proximity of bistable emission close to threshold and the existence
of ``dynamical traps'' via the folding of some unstable limit cycles
induced by XPR was already established in \cite{MGJ-PRA-07} as some
likely mechanisms for the degradation of the square wave signal, see
Fig.9 in \cite{MGJ-PRA-07} for instance. In our case, the system
does not exhibit such bistability since we are far from threshold.
Yet the proximity in parameter space of the line $\gamma_{a}+\alpha\gamma_{p}=0$
depicted in Fig.~\ref{fig:theo_0b} that interchange the stability
of the $X$ and $Y$ solution can play a similar role. We describe
in Fig.~\ref{fig:theo_6a} such degraded square wave dynamics in
the proximity of the parameter value $\gamma_{a}=-\alpha\gamma_{p}$
where the $X$ and $Y$ polarizations exchange their stability. Here
also we integrated the phase model given by Eqs.~(\ref{eq:Sig},\ref{eq:Phi})
and reconstructed the intensity of the Y component as $I_{y}\sim\left|1-e^{i\Phi}\right|^{2}$.
In Fig.~\ref{fig:theo_6a}, the dynamics experiences a critical slowing
down at the end of the second plateau where the LP-y component is
off and the escape from the weak saddle can be imagined as a noise
induced wandering in an almost flat landscape. 

\begin{figure}[tp]
\begin{centering}
\includegraphics[bb=0bp 0bp 552bp 333bp,width=0.49\textwidth]{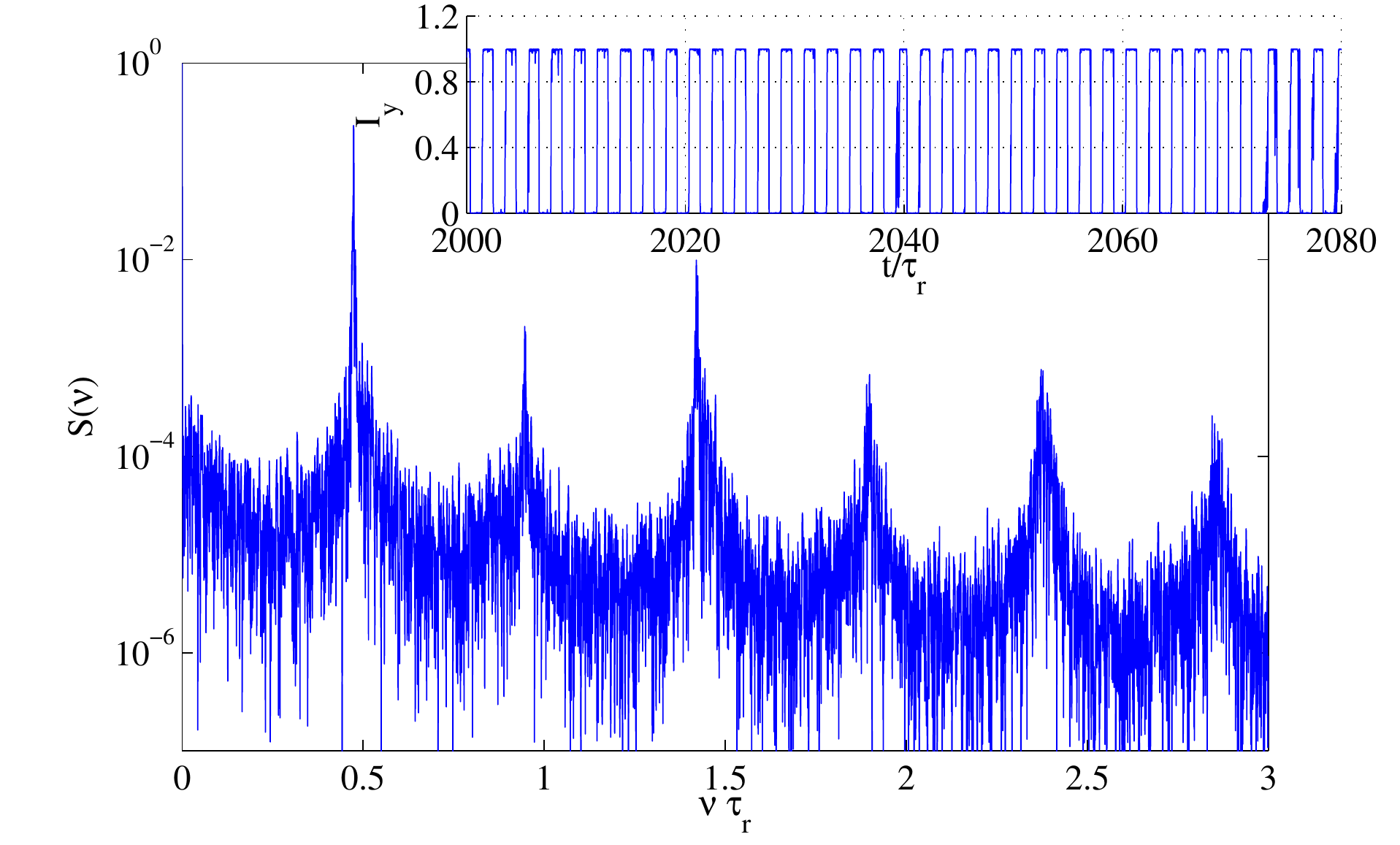}
\par\end{centering}

\centering{}\caption{(Color Online): Square wave switching dynamics (inset) and power spectrum
for $\eta=0.05$, $\gamma_{a}=-0.09$, $\beta=0.25$, $\tau_{r}=500$
and $\tau_{f}=1020$. The signal is much more regular, the period
much closer than $2\tau_{r}$. The power spectrum shows the signature
of a square wave signal with only odd harmonics. \label{fig:theo_6b} }
\end{figure}

However, a small amount of optical feedback has the effect to re-stabilize
the $Y$ polarization and incidentally to accelerate the escape of
the saddle represented by emission into the ÑLP-x mode. We describe
in Fig.~\ref{fig:theo_6b} such regime and show that even in the
proximity of polarization switching, robust square wave switching
can be obtained for proper choice of the feedback delay, i.e. $\tau_{r}\sim2\tau_{r}$.

\section{Conclusions}

In this manuscript we reduced the dynamics of the VCSEL far from the
laser threshold to a model that consists in two phases: the orientation
phase of the quasi-linear polarization and the optical phase of the
field. We showed that the dynamics remains confined close to the equatorial
plane of a Stokes sphere of a given radius which allowed us to decouple
the relaxation oscillation for the total emitted power as well as
the fluctuations in the ellipticity of the emitted light. 

Such simplification allowed expressing analytically the modes in presence
of XPR and PSF and to shed some light on the complex modal structure
given by the double feedback configuration. We also reinterpreted
the square waves switching dynamics \cite{JMB-PRL-06,MJB-JQE-07,SGP-PRA-2012,MJB-PRA-14}
previously found as polarization orientation kinks. Close to the polarization
switching the stability of both the LP-y and LP-x modes becomes marginal
which was shown to have a profound impact on the regularity of the
anti-phase square wave switching induced by XPR. We showed that also
in the reduced phase model the inclusion of optical feedback with
a proper delay can have the effect to regularize the dynamics and
that it can be used to mitigate the polarization degeneracy.

As future perspectives we believe that a similar approach can be applied
to the case of isotropic rotated feedback. Such effect was shown to
give anti-phase polarization oscillations \cite{LHG-APL-98} up to
frequencies $\sim10\,$GHz. Our method would yield a similar phase
model which would allow studying such polarization dynamics possibly
as a single dynamical equation. Also of great interest, our method
can be extended readily to the case of polarized, and possibly detuned,
optical injection.

\section*{Acknowledgments}

We acknowledge fruitful discussions with Salvador Balle. J.J. acknowledges
financial support from the Ramón y Cajal fellowship, the CNRS for
supporting a stay at the INLN where part of this work was developed
as well as financial support from project RANGER (TEC2012-38864-C03-01)
and from the Direcció General de Recerca, Desenvolupament Tecnològic
i Innovació de la Conselleria d'Innovació, Interior i Justí cia del
Govern de les Illes Balears co-funded by the European Union FEDER
funds. M.M. and M.G. acknowledge funding of Région \textquotedbl{}Provence-Alpes-Cote
d'Azur\textquotedbl{} with the \textquotedbl{}Projet Volet Générale
2011 : Génération et Détection des Impulsions Ultra Rapides (GEDEPULSE)\textquotedbl{}.

\bibliographystyle{unsrt}
\bibliography{/home/javaloye/Dropbox/BIBLIO/full,/home/javaloye/Dropbox/BIBLIO/javaloyes_articles}

\end{document}